\documentclass[aps,prc,a4paper,nofootinbib,showpacs,showkeys,
preprintnumbers,superscriptaddress,twocolumn]{revtex4}

\usepackage{amsmath}
\usepackage{amssymb}
\usepackage{bm}
\usepackage{graphicx}
\usepackage{color}

\begin{document}

\title{Initial state fluctuations from mid-peripheral to ultra-central collisions in a event-by-event transport approach}

\author{S. Plumari}
\affiliation{Department of Physics and Astronomy, University of Catania, Via S. Sofia 64, I-95125 Catania}
\affiliation{INFN-Laboratori Nazionali del Sud, Via S. Sofia 62, I-95123 Catania, Italy}

\author{G. L. Guardo}
\affiliation{Department of Physics and Astronomy, University of Catania, Via S. Sofia 64, I-95125 Catania}
\affiliation{INFN-Laboratori Nazionali del Sud, Via S. Sofia 62, I-95123 Catania, Italy}

\author{F. Scardina}
\affiliation{Department of Physics and Astronomy, University of Catania, Via S. Sofia 64, I-95125 Catania}
\affiliation{INFN-Laboratori Nazionali del Sud, Via S. Sofia 62, I-95123 Catania, Italy}

\author{V. Greco}
\affiliation{Department of Physics and Astronomy, University of Catania, Via S. Sofia 64, I-95125 Catania}
\affiliation{INFN-Laboratori Nazionali del Sud, Via S. Sofia 62, I-95123 Catania, Italy}

\begin{abstract}
We have developed a relativistic kinetic transport approach that incorporates initial state fluctuations 
allowing to study the build up of elliptic flow $v_2$ and high order harmonics $v_3$, $v_4$ and $v_5$ for a 
fluid at fixed $\eta/s(T)$.
We study the effect of the $\eta/s$ ratio and its T dependence on the build up of the $v_n(p_T)$
for two different beam energies: RHIC for Au+Au at $\sqrt{s}=200 \,GeV$ 
and LHC for $Pb+Pb$ at $\sqrt{s}=2.76 \,TeV$. We find that for the two different beam energies 
considered the suppression of the $v_n(p_T)$ due to the viscosity of the medium have different contributions 
coming from the cross over or QGP phase. 
Our study reveals that only in ultra-central collisions ($0 - 0.2 \%$) the $v_n(p_T)$ have a stronger sensitivity 
to the T dependence of $\eta/s$ in the QGP phase and this sensitivity increases with the order of the harmonic n. 
Moreover, the study of the correlations between the initial spatial anisotropies $\epsilon_n$ and the final flow 
coefficients $v_n$ shows that at LHC energies there is more correlation than at RHIC energies. The degree of 
correlation increases from peripheral to central collisions, but only in ultra-central collisions at LHC, 
we find that the linear correlation coefficient $C(n,n) \approx 1$ for $n=2,3,4$ and $5$. 
This suggests that the final correlations in the ($v_n$,$v_m$) space reflect the initial correlations in the 
($\epsilon_n$,$\epsilon_m$) space.
\end{abstract}


\maketitle

\section{Introduction}

The experimental results accumulated in these years in the ultra relativistic heavy ion
collisions (uRHICs) first in the experiments conducted at RHIC and more recently at the 
LHC has shown that the elliptic flow 
$v_2= \langle cos( 2\, \varphi_p) \rangle= \langle (p_x^2-p_y^2)/(p_x^2+p_y^2) \rangle$, 
is the largest ever observed in HIC \cite{STAR_PHENIX,Aamodt:2010pa}.
The elliptic flow is a measurement of the momentum anisotropy of the emitted particles and it is 
an observable that encodes information about the transport properties of the matter created in 
these collisions. 
Theoretical calculations within viscous hydrodynamics \cite{Romatschke:2007mq,Heinz} 
and in the recent years also calculation performed within transport approach 
\cite{Ferini:2008he,Xu:2008av,Plumari_Bari} have shown that 
this large value of $v_2$ is consistent with a matter with a very low shear viscosity to entropy density 
ratio $\eta/s$ close to the conjectured lower bound for a strongly interacting system, $\eta/s=1/4\pi$ \cite{Kovtun:2004de}.

While early studies have been focused on elliptic flow generated by the global 
almond shape of the fireball for non central collisions. 
In the recent years the possibility to measure experimentally the event-by-event 
angular distribution of emitted particle has made possible to go beyond such a simplified picture 
accessing the fluctuating shape that encodes higher order harmonics generating non zero flows 
$v_n= \langle cos(n\, \varphi_p) \rangle$ \cite{Adare:2011tg,Richardson:2012kq,ATLAS:2012at}. 
Hence most of the research activity has been now focused on the study of the effects of the 
fluctuations in the initial geometry due to the fluctuations of the position of 
the nucleons in the overlap region of the collision \cite{Petersen:2010cw,Qin:2010pf,Holopainen:2010gz,Schenke:2011bn,Gale:2012rq,Bravina:2013xla}. 
Such fluctuations in the initial 
geometry are sources for momentum anisotropies of any n-th order harmonics $v_n= \langle cos(n\, \varphi_p) \rangle$
and in particular of the triangular flow $n=3$, that especially in ultra-central collisions appears as the largest one \cite{Abelev:2012di,ATLAS:2012at,CMS:2013bza}.

The comparison between event-by-event viscous hydrodynamical calculations and the experimental results for $v_n$ 
seems to confirm a finite but not too large value of $4 \pi \eta/s \sim 1-3$ 
\cite{Schenke:2011bn,Gale:2012rq}.
However, small values of $\eta/s$ is not an evidence of the creation of a QGP 
phase. A phenomenological estimation of its temperature dependence could give 
information if the matter created in these collisions undergoes a phase transition 
\cite{Csernai:2006zz,Lacey:2006bc,Plumari:2013bga}.
Information about a temperature dependence of $\eta/s$ can be achieved studying the $v_2(p_T)$ 
and the high order harmonic $v_n(p_T)$ in a wider range of energies. Similar studies have been 
performed using a transport approach but only for the elliptic flow  in an approach not 
incorporating event-by-event fluctuations \cite{Plumari:2013bga,Plumari:2014rga}. In this paper we extend this analysis to high order harmonics 
studying the role of the $\eta/s$ on the build up of $v_n(p_T)$ using for the first time a cascade approach with 
initial state fluctuations.

\begin{figure}[h]
\begin{minipage}{16pc}
\includegraphics[width=16pc]{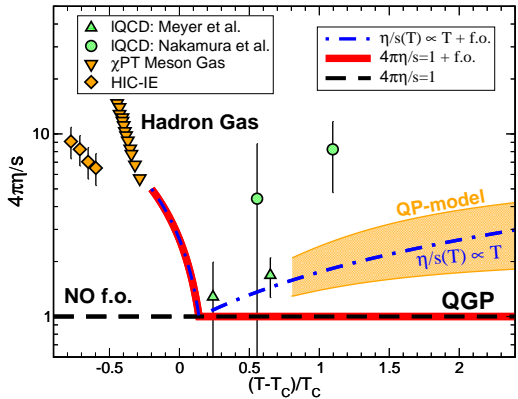}
\end{minipage}
\hspace{1pc}%
\begin{minipage}{20pc}\caption{
Different parametrizations for $\eta/s$ as a function of the temperature.
The orange area refers to the quasi-particle model predictions
for $\eta/s$ \cite{Plumari:2011mk}.
The three different lines indicate different possible T dependencies studied in this paper. 
Symbols are as in the legend.
See the text for more details.\label{Fig:etas_T}
}
\end{minipage}
\end{figure}

There are several theoretical indications that $\eta/s$ should 
have a particular behavior with the temperature \cite{Csernai:2006zz,Lacey:2006bc,Prakash:1993bt,Chen:2007xe,Meyer:2007ic,Das:2010yi}. 
As an example in Fig.\ref{Fig:etas_T} it is shown a collection of theoretical 
results about the temperature dependence of $\eta/s$. Fig.\ref{Fig:etas_T}
shows that in general $\eta/s$ should have a typical behavior of phase transition 
with a minimum close to the critical temperature $T_C$ 
\cite{Csernai:2006zz,Lacey:2006bc,Plumari:2013bga,Plumari:2014rga}.
On one hand at low temperature estimates of $\eta/s$ in the 
chiral perturbation theory for a meson gas \cite{Prakash:1993bt,Chen:2007xe}, 
have shown that in general $\eta/s$ is a decreasing function with the 
temperature, see down-triangles in Fig.\ref{Fig:etas_T}. Similar results 
for $\eta/s$ have been extrapolated from heavy-ion collisions at intermediate 
energies, see HIC-IE diamonds in Fig.\ref{Fig:etas_T}. 
On the other hand at higher temperature $T>T_c$ lQCD calculation have 
shown that in general $\eta/s$ becomes an 
increasing function with the temperature \cite{Meyer:2007ic,Nakamura:2004sy}, see up-triangles 
and circles in Fig.\ref{Fig:etas_T}, but due to the large error bars in the lQCD results for $\eta/s$ 
it is not possible to infer a clear temperature dependence in the QGP phase. 
The analysis at different energies of $v_2(p_T)$ and the extension to high order 
harmonics $v_n(p_T)$ can give further information about the T dependence 
of $\eta/s$. 
In this paper we study and discuss the build-up of anisotropic flows $v_n$ in 
ultra-relativistic HIC treating the system as a fluid with some $\eta/s(T)$. 
This is achieved by mean of a transport approach with initial state fluctuations.
The paper is organized as follows. 
In Section \ref{section:approach}, we introduce the transport approach at fixed shear viscosity 
to entropy density $\eta/s$.
In Section \ref{section:initial}, we discuss the initial conditions and in particular the 
implementation of the initial state fluctuations in the transport approach.
In Section \ref{section:etas}, we study the time evolution of the anisotropic flows $\langle v_n \rangle$
and the effect of the $\eta/s(T)$ on the differential $v_n(p_T)$.
Finally in Section \ref{section:corr} we study the correlations between the initial asymmetry in 
coordinate space measured by the coefficients $\epsilon_n$ and the final anisotropy in momentum 
space measured by the anisotropic flows $\langle v_n \rangle$.
In this paper we will show results on $v_n(p_T)$ for $n=2,3,4$ and $5$ for the two different 
systems $Au+Au$ at $\sqrt{s}=200 \, GeV$ and $Pb+Pb$ at $\sqrt{s}=2.76 \, TeV$ at different centralities.

\section{Kinetic approach at fixed shear viscosity to entropy density ratio}\label{section:approach}
In this work we employ the kinetic transport theory to study the evolution of the fireball 
created in relativistic heavy-ion collisions.   
We perform such simulations using a relativistic 
transport code developed in these years to perform studies of the dynamics 
of heavy-ion collisions at both RHIC and LHC energies \cite{Ferini:2008he,Plumari_Bari,Plumari:2012xz,Plumari:2012ep,Ruggieri:2013bda,Ruggieri:2013ova}.
The evolution of the phase-space distribution function $f(x, p, t)$ is given by solving the Relativistic Boltzmann Transport (RBT) equation:
\begin{eqnarray}
\label{RBT}
& & p^{\mu}\, \partial_{\mu} f(x,p)= C[f] + S[f_0]
\end{eqnarray}
where $C[f]$ is the Boltzmann-like collision integral. In the result shown in this paper we 
have considered only the $2 \leftrightarrow 2$ processes and for one component system $C[f]$ can be written as,
\begin{eqnarray}
C[f]=\int_{2,1^\prime,2^\prime} (f_{1^\prime}f_{2^\prime} - f f_2) |{\cal M}|^2\delta^4(p + p_2 - p_{1^\prime} - p_{2^\prime})~
\end{eqnarray}

where $\int_{2,1^\prime,2^\prime}=\int \Pi_{k=2,1^\prime,2^\prime} d^3 {p}_{k}/ 2 E_{k} (2\pi)^3 $ and ${\cal M}$ denotes the transition amplitude for the elastic processes which is 
directly linked to the differential cross section $|{\cal M}|^2=16 \pi\,s\,(s-4M^2) d\sigma/dt$ with $s$ the Mandelstam invariant.
Numerically we solve the RBT equation using the so called test particle method and the collision integral is solved by using Monte Carlo 
methods based on the stochastic interpretation of transition amplitude \cite{Xu:2004mz,Ferini:2008he,Plumari:2012ep}.

In the standard use of the transport theory one fixes the microscopical details of the scattering 
like matrix element or cross sections of the processes to study 
the effect of the microscopical details on the observables. 
This is however not our aim we exploit the cross section $\sigma_{tot}$ as a tool to determine 
the $\eta/s$ of the system. As shown in \cite{Plumari:2015sia} in the hydrodynamic limit observables like 
$v_2(p_T)$ or spectra don't depend on the microscopic details encoded in $|{\cal M}|^2$.
In agreement with the implicit assumption of hydrodynamics.
In such an approach it is possible to study directly the impact of $\eta/s$ on observables 
like the anisotropic flows $v_n(p_T)$ which is the main focus of this paper.
Compared with the viscous hydrodynamic calculations a kinetic approach at fixed 
$\eta/s$ has manly two advantages: first in this approach we start from a description in terms of $f(x,p)$
instead of starting from $T^{\mu \nu}(x)$ and it is possible to include initial non equilibrium effects 
(see \cite{Ruggieri:2013bda,Ruggieri:2013ova}). 
Second, this approach is not based on an ansatz for the viscous corrections for the phase-space distribution 
function $\delta f$ with the limitation in the transverse momentum range in order to ensure that $\delta f/f << 1$. 
Also this approach provides a tool to study the effect of $\eta/s$ on the observables in a wider range of 
$\eta/s$ and in transverse momentum $p_T$. 
Notice also that the kinetic freeze-out can be determined self-consistently with an increasing $\eta/s(T)$ that determines 
a smooth switching-off of the scattering rates.
A more detailed discussion can be found in previous papers, see \cite{Ferini:2008he,Ruggieri:2013bda,Ruggieri:2013ova}. 
The disadvantage of the present approach is that hadronization has not yet been included. A more general 
disadvantage is that RBT converge to viscous hydrodynamics with the relaxation time typical of a kinetic theory.
However viscous hydrodynamics with relaxation times of kinetic theory have been shown to be in quite good agreement with experimental 
data.

In order to study the dynamical evolution of the fireball with a certain 
$\eta/s(T)$ we determine locally in space and time the total cross section 
$\sigma_{tot}$ needed to have the wanted local viscosity.
As shown in \cite{Plumari:2012ep} the Chapmann-Enskog theory correctly describes the relation 
between $\eta \leftrightarrow T, \sigma(\theta), \rho$ providing a good agreement with the results 
obtained using the Green-Kubo correlator.
In the Chapmann-Enskog theory and for a pQCD inspired cross section, typically used in parton cascade approaches 
\cite{Zhang:1999rs,moln02,Ferini:2008he,Greco:2008fs,Plumari_njl,Xu:2004mz,Xu:2008av}, $d\sigma/dt \sim \alpha_s^2/(t-m_D^2)^2$,
the $\eta/s$ is given by the following expression:
\begin{equation}
\eta/s =\frac{1}{15} \langle p\rangle \, \tau_{\eta}=
\frac{1}{15}\frac{ \langle p\rangle}{ g(a) \sigma_{tot} \rho} \,,
\label{eq:etas_CE}
\end{equation}
where $a=m_D/2T$, with $ m_D$ being the screening mass regulating 
the angular dependence of the cross section, while $g(a)$ is the 
proper function accounting for the pertinent relaxation time 
$\tau_{\eta}^{-1}=g(a) \sigma_{tot} \rho$ associated
to the shear transport coefficient and it is given by:
\begin{eqnarray}
g(a)=\frac{1}{50}\! \int\!\! dyy^6
\left[ (y^2{+}\frac{1}{3})K_3(2 y){-}yK_2(2y)\right]\!
h\left(\frac{a^2}{y^2}\right),
\label{g_CE}
\end{eqnarray}
where $K_n$-s are the Bessel functions and the function 
$h$ relate the transport cross section to the total cross section
$\sigma_{tr}(s)= \sigma_{tot} \, h(m_{D}^2/s)$
with $h(\zeta)=4 \zeta ( 1 + \zeta ) \big[ (2 \zeta + 1) ln(1 + 1/\zeta) - 2 \big ]$.

In order to study the role of the $\eta/s$ ratio and its temperature dependence 
we consider three different cases: one with a constant 
$4\pi \eta/s=1$ during all the evolution of the system dashed line in Fig.\ref{Fig:etas_T}
another one with $4\pi \eta/s=1$ at higher temperature in the QGP phase 
and an increasing $\eta/s$ in the cross over region towards the estimated value 
for hadronic matter $4\pi \eta/s \approx 6$ \cite{Chen:2007xe,Demir:2008tr} shown by solid line in Fig.\ref{Fig:etas_T}.
Such an increase of $\eta/s$ in the cross over region $0.8 T_C \le T \le 1.2 T_C$ allows 
for a smooth realistic realization of the kinetic freeze-out. 
This is because at lower temperature, according to the formula 
Eq.(\ref{eq:etas_CE}) $\sigma \propto (\eta/s)^{-1}$ i.e. the increase 
of $\eta/s$ towards the estimated value for the hadronic matter implies the total cross section 
decrease and this permits to achieve in a self-consistent way the kinetic freeze-out.
In the following discussion the term f.o. means to take into account the increase of 
$\eta/s$ at low temperature. The third one is shown in Fig.\ref{Fig:etas_T} by the dot 
dashed line. In this case we consider the increase of $\eta/s$ at higher temperature 
with a linear temperature dependence and a minimum close to the critical temperature
with a temperature dependence similar to that expected from general considerations
as shown in Fig.\ref{Fig:etas_T}.

\section{Initial conditions}\label{section:initial}
\begin{figure}
\begin{center}
\includegraphics[width=10pc]{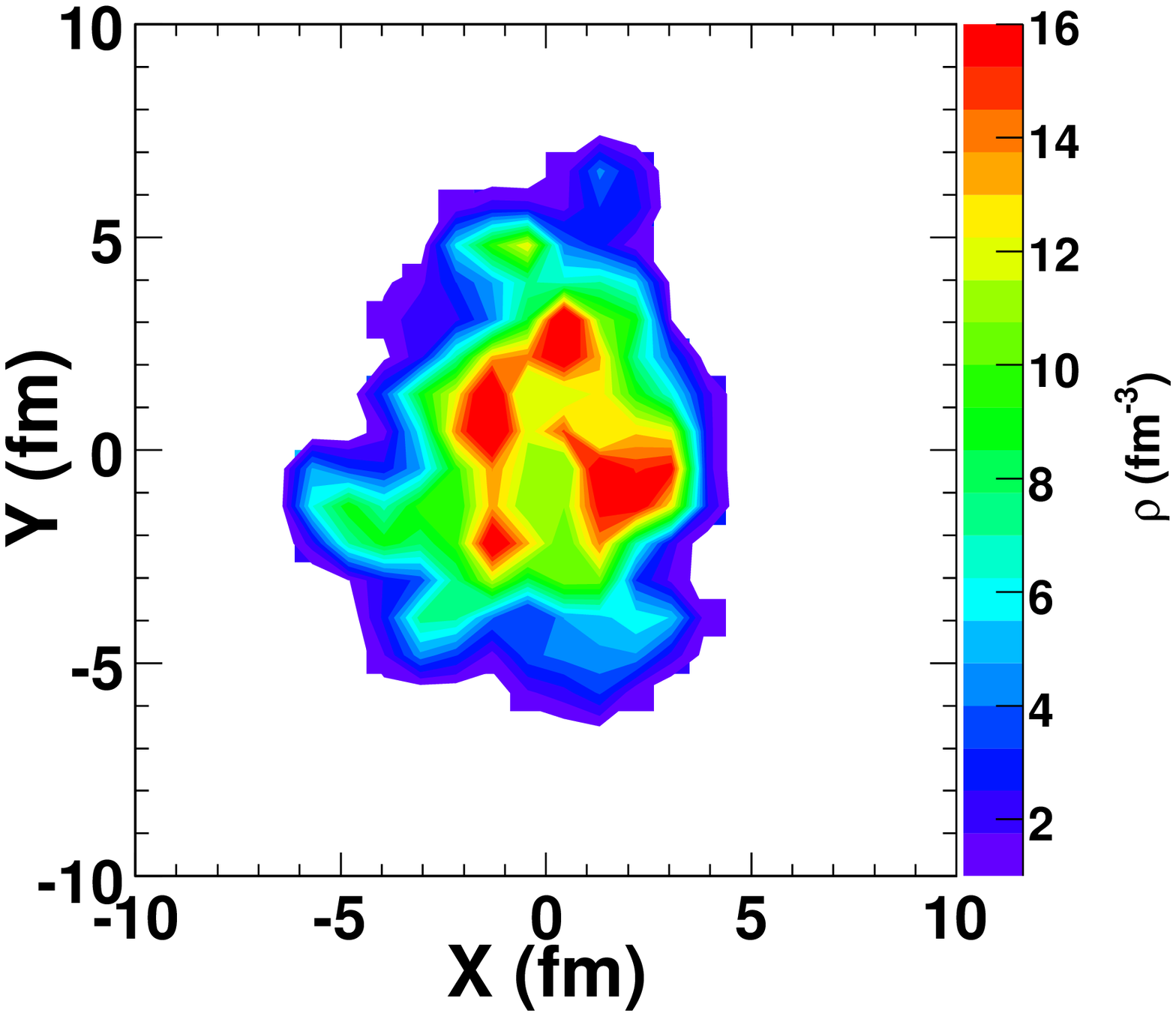}
\includegraphics[width=10pc]{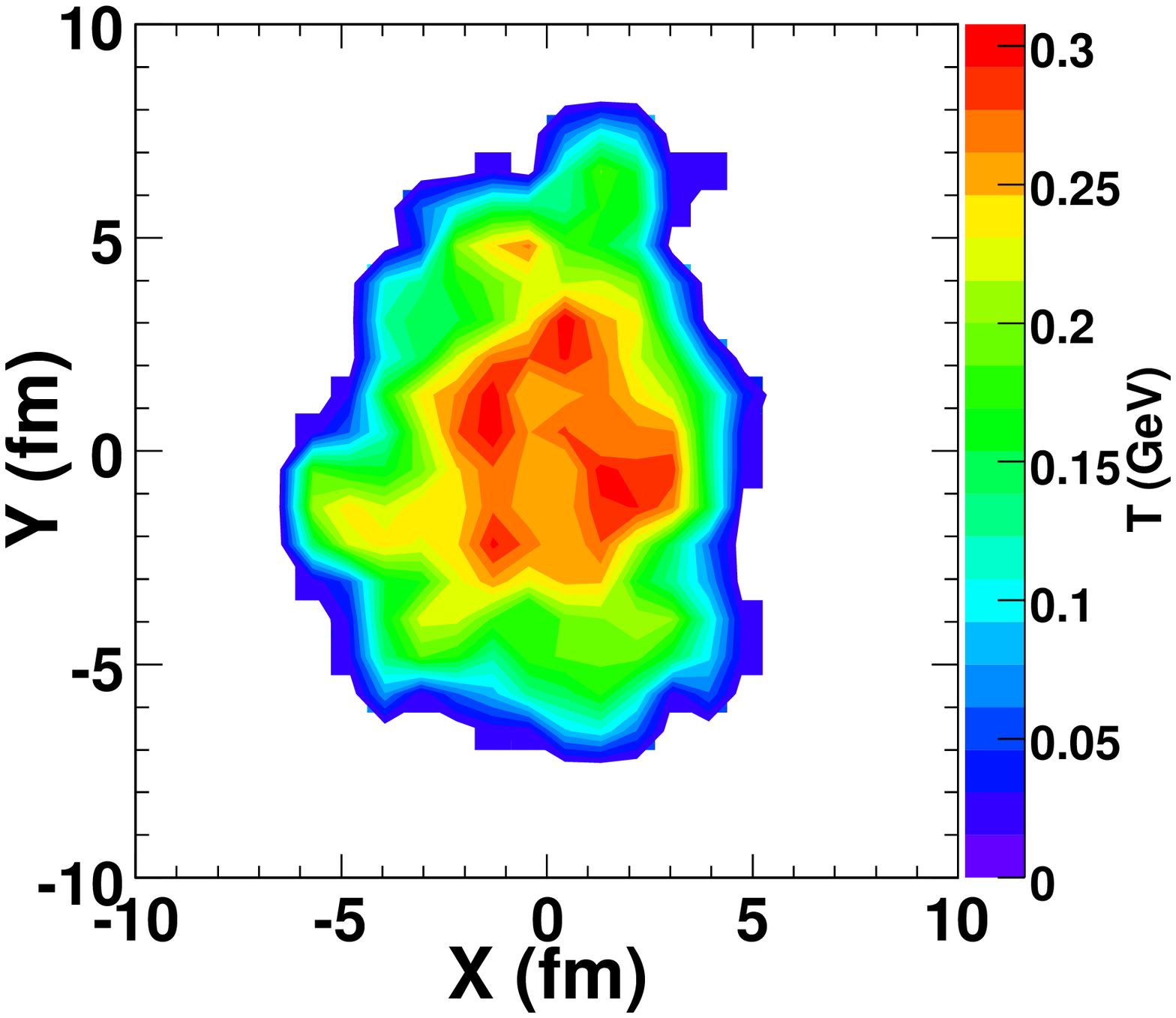}
\hspace{1pc}
\includegraphics[width=10pc]{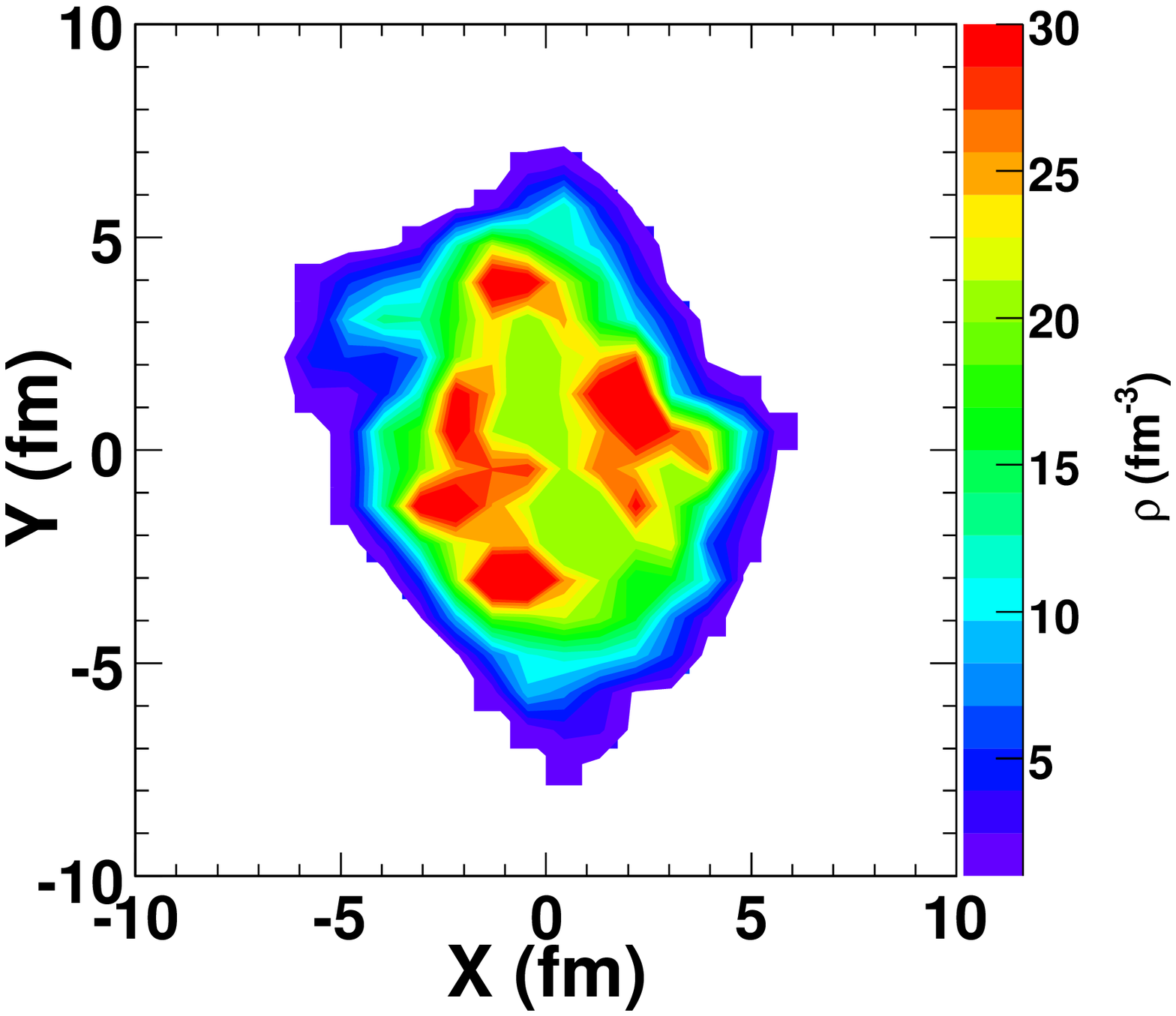}
\includegraphics[width=10pc]{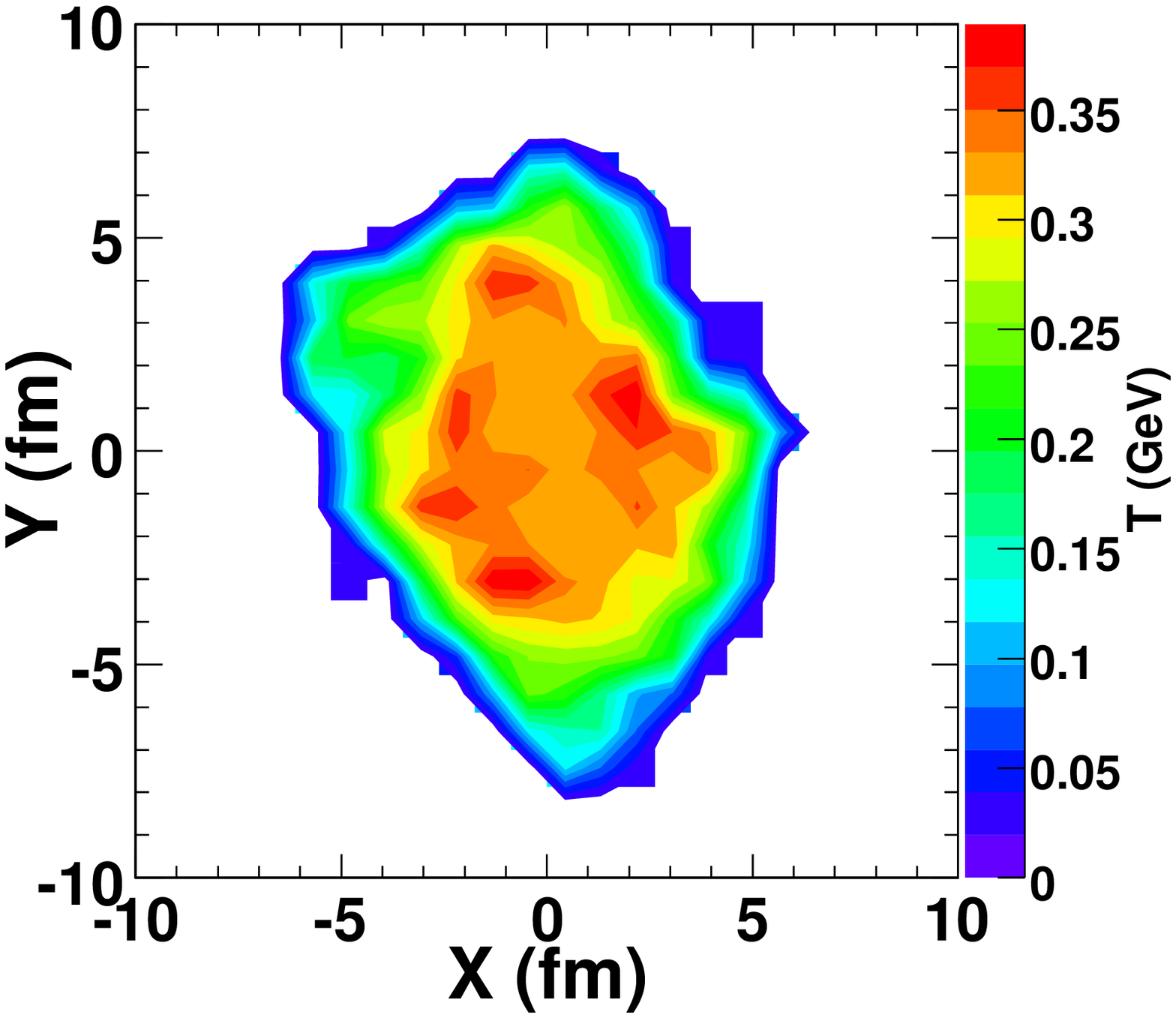}
\end{center}
\caption{In the left column it is shown the initial transverse density $\rho_T(x,y)$ at mid rapidity for two typical events for $Au+Au$ at $\sqrt{s_{NN}}=200 \, GeV$ (upper panel) and $Pb+Pb$ at $\sqrt{s_{NN}}=2.76 \, TeV$ (lower panel). In the right column it is shown the corresponding initial temperature in transverse plane. These plots are for an impact parameter of $b=7.5 \, fm$.
\label{Fig:rho_T_profile}}
\end{figure}

The main novelty in the present paper is the implementation of initial state fluctuations
in a transport cascade approach.
We will consider two systems at different centralities:
$Au+Au$ collisions at $\sqrt{s_{NN}}=200 \, GeV$ produced at RHIC and $Pb+Pb$ 
collisions at $\sqrt{s_{NN}}=2.76 \, TeV$ at LHC. In particular in this section 
we discuss the implementation of the initial state fluctuations in the above
transport approach.
In order to generate an event by event initial profile 
we use the Monte-Carlo Glauber model.
In this model the Woods-Saxon distribution is used to sample randomly the positions 
of the nucleons in the two colliding nucleus $A$ and $B$. In this way a discrete 
distribution for these nucleons is generated. 
We employ the geometrical method to determine if the two nucleons one from the nucleus 
$A$ and the other one from the nucleus $B$ are colliding. 
Within this method two nucleons collide each other if the relative distance in the 
transverse plane is $d_{T} \leq \sqrt{\sigma_{NN}/\pi}$ where $\sigma_{NN}$ 
is the nucleon-nucleon cross section. In our calculation we have used 
$\sigma_{NN}=4.2 \, fm^2$ for RHIC and $\sigma_{NN}=7.0 \, fm^2$ for LHC. 
$N_{coll}$ and $N_{part}$ are given by counting the number of 
collisions and the number of participating nucleons for each event.
The next step is the conversion of the discrete distribution for the 
nucleons into a smooth one by assuming for each nucleon a gaussian 
distribution centered in the nucleon position.
In our model we choose to convert the information of the nucleon distribution 
into the density in the transverse plane $\rho_T(x,y)$ which is given by the following sum
\begin{eqnarray}
\rho_T(x,y) = C \sum_{i=1}^{N_{part}} \exp{\bigg[-\frac{(x-x_i)^2+(y-y_i)^2}{2\sigma_{xy}^2}\bigg]}
\label{Eq:rho_T}
\end{eqnarray}
where $C$ is an overall normalization factor fixed by the longitudinal distribution $dN/dy$ while 
$\sigma_{xy}$ is the Gaussian width which regulates the smearing of the fluctuations and in the 
following calculations it has been fixed to $\sigma_{xy} = 0.5 \, fm$.
In our calculation we have assumed initially a longitudinal boost invariant distribution from $y=-2.5$ to $y=2.5$.
In the first column of Fig.\ref{Fig:rho_T_profile} it is shown the contour plot of the initial transverse 
density at mid rapidity for a given event with impact parameter $b=7.5 \, fm$. The upper panel refers to 
the system $Au+Au$ at $\sqrt{s_{NN}}=200 \, GeV$ and the lower panel to $Pb+Pb$ at $\sqrt{s_{NN}}=2.76 \, TeV$.

\begin{figure}
\begin{center}
\includegraphics[width=18pc]{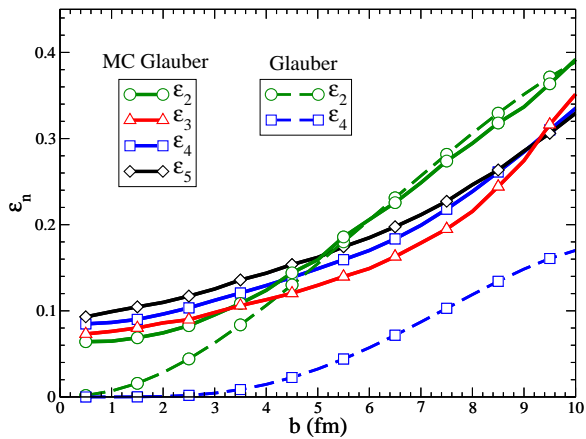}
\end{center}
\caption{Initial spacial anisotropies $\epsilon_n$ as a function of the impact parameter. 
Different symbols are for different $n$. The solid lines refer to the Monte Carlo Glauber while 
the dashed ones to the optical Glauber model. 
\label{Fig:eps_n}}
\end{figure}

The transverse density $\rho_T(x,y)$ fixes the initial anisotropy 
in coordinate space that is quantified in 
terms of the following coefficients $\epsilon_n$:
\begin{equation}
\epsilon_{n}=\frac{\sqrt{\langle r_T^n \cos{(n \phi)} \rangle^2 + \langle r_T^n \sin{(n \phi)} \rangle^2}}{\langle r_T^n\rangle}
\label{Eq:eps_n}
\end{equation}
where $r_T=\sqrt{x^2+y^2}$ and $\phi=arctan(y/x)$ is the polar coordinate in the transverse plane.
In Fig.\ref{Fig:eps_n} it is shown the initial spatial anisotropies $\epsilon_2, \, \epsilon_3, \, \epsilon_4$
and $\epsilon_5$ as a function of the impact parameter. The second coefficient $\epsilon_2$ shows a stronger 
dependence with the impact parameter with respect to the other coefficients 
because it acquires a contribution due to the global almond shape of the fireball while the other 
harmonics have most of their origin in the fluctuations of the positions of the nucleons. 
For more central collisions $b \le 2.5 \, fm$ the $\epsilon_2$ is even smaller than the other harmonics 
because when the effect of the elliptic overlap region disappears it becomes more difficult to have fluctuations of the positions 
of the nucleons along only one preferential direction.

For the initialization in momentum space at RHIC (LHC) energies we 
have considered for partons with transverse momentum $p_T \leq p_0=2 \, GeV$ 
($3 \, GeV$) a thermalized spectrum in the transverse plane.
Assuming the local equilibrium the initial local temperature in the transverse plane $T(x,y)$ 
is evaluated by using the standard thermodynamical relation $\rho_T(x,y)=\gamma T^3/\pi^{2}$
with $\gamma=2 \times (N_c^2-1) + 2 \times 2 \times N_c \times N_f=40$ with $N_c=3$ and $N_f=2$.
In the right column of Fig.\ref{Fig:rho_T_profile} it is shown the 
corresponding initial local temperature in transverse plane. 
As shown in the central region of the fireball for mid peripheral collision we can reach 
temperature $T \approx 300 \, MeV$ at RHIC and $T \approx 400 \, MeV$ at LHC.
While for partons with $p_T > p_0$ we have assumed the spectrum 
of non-quenched minijets according to standard NLO-pQCD calculations with a power law shape \cite{Greco:2003xt,Greco:2003mm}. 
In coordinate space the partons with $p_T > p_0$ have been distribuited according the binary collisions.
The initial transverse momentum of the particles is distributed uniformly in the azimuthal angle.
We fix the initial time of the simulation to $\tau_0=0.6 \, fm/c$ for RHIC and 
$\tau_0=0.3 \, fm/c$ for LHC. 

In the following discussion, we will consider two 
different types of initial conditions.
One consisting in a fixed initial distribution 
by using the standard Glauber model as used in previous works, see 
\cite{Ferini:2008he,Plumari_Bari,Plumari:2012xz,Plumari:2013bga}. 
The second one consisting of an initial profile changing event by event
according to the MC Glauber model as discussed before.

In our simulations we have used $N_{event}=500$ events for each centrality class.
This number is enough to get solid results for the spectra, differential elliptic flow and high 
order flow coefficients $v_n(p_t)$. For the study of the correlations between the initial 
$\epsilon_n$ and the final $v_n$ that will be shown in the next section we have extended this analysis 
to $10^{3}$ events. 
The inclusion of the initial state fluctuations introduce a further difficulties because 
in order to get stable results we need to have a good sampling of the initial geometry event 
by event and this is controlled by the total number of test particles $N_{test}$. 
Furthermore an irregular initial profile need a good calculation grid resolution.
We have checked the convergency of our results for $v_2$, $v_3$ and $v_4$ with the 
lattice spacing of the calculation grid and $N_{test}$. We found the convergency for a grid with a transverse area of the cell  
$A_T=0.12 fm^2$ and $N_{test}=2 \cdot 10^6$ as total number of test particles per event.

\begin{figure}
\begin{center}
\includegraphics[width=20pc]{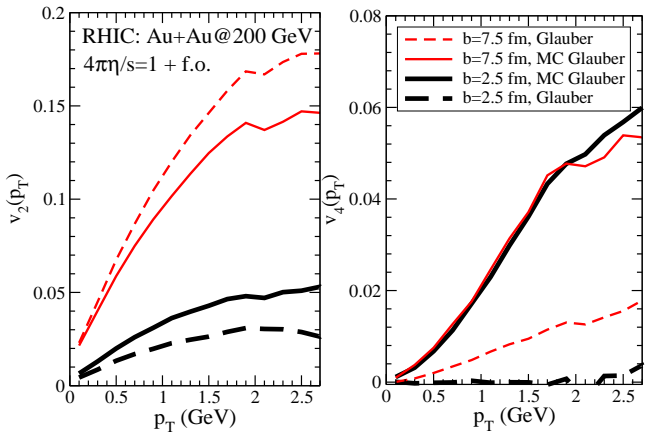}
\end{center}
\caption{Results for $Au+Au$ collisions at $\sqrt{s_{NN}}=200 \, GeV$ for mid rapidity.
Left: differential elliptic flow $v_2(p_T)$ at mid rapidity. The solid lines refer to the case with initial 
state fluctuations the dashed lines are for the case without initial fluctuations.
Thick lines are for $b=2.5 \, fm$ while thin lines for $b=7.5 \, fm$.
Right: differential $v_4(p_T)$ at mid rapidity with the same legend as in the left panel.
\label{Fig:v24}}
\end{figure}

The elliptic flow $v_2(p_T)$ and the high order harmonics $v_3(p_t)$, $v_4(p_T)$ and $v_5(p_T)$
have been calculated as
\begin{equation}
v_n=\langle \cos{[n(\phi -\Psi_n)]}\rangle
\label{Eq:v_n}
\end{equation}
where the momentum space angles $\Psi_n$ are given by
\begin{equation}
\Psi_n=\frac{1}{n}\arctan{\frac{\langle\sin{(n\phi)}\rangle}{\langle\cos{(n\phi)}\rangle}}
\end{equation}

In this section first we discuss the comparison between the Glauber model w/o initial fluctuations with the MC Glauber 
with fluctuations. Without initial state fluctuations only even harmonics can be generated therefore we will consider here only $v_2(p_T)$
 and $v_4(p_T)$.
In the left panel of Fig.\ref{Fig:v24} we compare the differential elliptic 
flow $v_2(p_T)$ obtained with an initial state that changes event by event according to the MC 
Glauber model (solid lines) as discussed in details the previous section with the one obtained 
for the case with an averaged initial profile (dashed lines). 
These results are for $Au+Au$ collisions at $\sqrt{s}=200 \, GeV$ and for $20-30 \%$ 
centrality class. In these calculations we have considered $4\pi \eta/s=1$ at high temperature 
and an increasing $\eta/s$ at lower temperature as shown by the red solid line in Fig.\ref{Fig:etas_T}.
As shown for mid peripheral collision (with $b=7.5 \, fm$) the effect of the fluctuations in the initial geometry 
is to reduce the $v_2(p_T)$ of about $15\%$, despite the same initial eccentricity $\epsilon_2$ in Glauber and MC Glauber, see green solid and dashed 
lines in Fig.\ref{Fig:eps_n}.
The reduced efficiency in building up  the $v_2(p_T)$ is related to the fact that for an irregular geometry in the transverse 
plane the pressure gradients generate also a small counter-flow towards the inner part of the 
fireball reducing the azimuthal anisotropy in momentum space due to the global almond shape.  
The introduction of the fluctuations in the initial geometry 
play the role to generate the higher order harmonics in particular the 
odd harmonics which were absent by symmetry in the averaged initial 
configuration.
In the right panel of Fig.\ref{Fig:v24} we show the same comparison 
for the quadrangular flow $v_4(p_T)$. We observe an opposite behaviour: 
the initial state fluctuations increase the final $v_4(p_T)$ by a factor of $3$.
This result is related to the fact that the fluctuations introduce 
about a factor $3$ larger initial $\epsilon_4$ as shown by the comparison between blue solid and dashed 
lines in Fig.\ref{Fig:eps_n}.
In other words for mid peripheral collisions most of $v_2(p_T)$ comes from the global 
almond shape while $v_4(p_T)$ comes normally from the initial fluctuations.
In fact as shown by the black thick solid and dashed lines in the left panel of 
Fig.\ref{Fig:eps_n} the effect of the fluctuations is to produce a larger $v_2(p_T)$. 
From the comparison between thick black solid and dashed line in the right panel of Fig.\ref{Fig:eps_n} 
we observe a non zero $v_4(p_T)$ that was absent in the averaged initial profile where 
the initial $\epsilon_4 \approx 0$ (see blue dashed line in Fig.\ref{Fig:eps_n}). 
Moreover we observe a low sensitivity of $v_4(p_T) $ with the centrality similarly to the 
experimental data at RHIC energies \cite{Richardson:2012kq}. Such a behaviour would be impossible to 
explain without initial state fluctuations.
\begin{figure}
\begin{center}
\includegraphics[width=18pc]{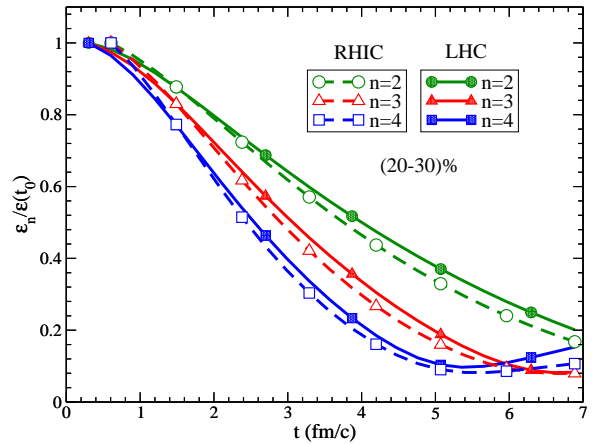}
\end{center}
\caption{$\epsilon_n/\epsilon_n(t_0)$ as a function of the time for $Au+Au$ collisions at $\sqrt{s_{NN}}=200 \, GeV$ (dashed lines) and for $Pb+Pb$ collisions at $\sqrt{s_{NN}}=2.76 \, TeV$ (solid lines). Different symbols refer to different harmonics $n$.
\label{Fig:eccn_time}}
\end{figure}

\begin{figure}
\begin{center}
\includegraphics[width=18pc]{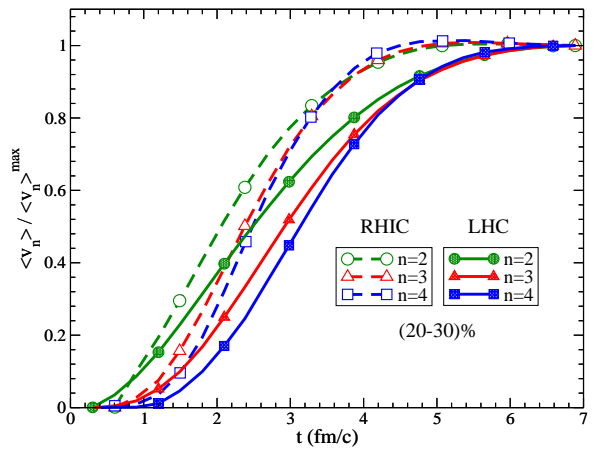}
\end{center}
\caption{$\langle v_n\rangle / \langle v_n\rangle^{max}$ as a function of time at mid rapidity and for $(20-30) \%$ of centrality.
Dashed lines are the results for $Au+Au$ at $\sqrt{s_{NN}}=200 \, GeV$ while solid lines $Pb+Pb$ at $\sqrt{s_{NN}}=2.76 \, TeV$.
Different symbols correspond to different harmonics. 
\label{Fig:vn_t}}
\end{figure}

\section{Effects of $\eta/s(T)$ on the $v_n(p_T)$}\label{section:etas}
In the first part of this section we discuss the time evolution of the eccentricities $\epsilon_n$ 
and the anisotropic flows coefficients $\langle v_n\rangle$ for $Au+Au$ collisions at 
$\sqrt{s}=200 \, GeV$ (solid lines) and for $Pb+Pb$ collisions at $\sqrt{s}=2.76 \, TeV$
In Fig.\ref{Fig:eccn_time} we plot the time evolution of the $\epsilon_n$ normalized 
to the initial eccentricity $\epsilon_n(t_0=0.6 \, fm/c)$ for RHIC and $\epsilon_n(t_0=0.3 \, fm/c)$ for LHC.
At very early times the small deformation of the fireball in the transverse plane decrease linearly with time and 
at first order of this deformation we have that $\epsilon_n \propto \epsilon_n(t_0)-\alpha_n \, t^{n-2}$.
This gives the ordering in the time evolution of $\epsilon_n$ shown in Fig.\ref{Fig:eccn_time}.
the time evolution of $\epsilon_n$ is faster for larger $n$. 

On contrary $v_n$ show an opposite behaviour during the early times of the expansion of the 
fireball. In Fig.\ref{Fig:vn_t} it shown the average $\langle v_n\rangle$ normalized
to its maximum value at the end of the expansion. The $\langle v_n\rangle$
appear later for larger $n$ and their development is flatter at early times for larger $n$. Similar results have 
been obtained in a $2+1D$ transport approach where considerations on the early times evolution of the fireball 
give that $\langle v_n\rangle \propto t^n$ \cite{Alver:2010dn, Gombeaud:2007ub}.
\begin{figure}
\begin{center}
\includegraphics[width=19pc]{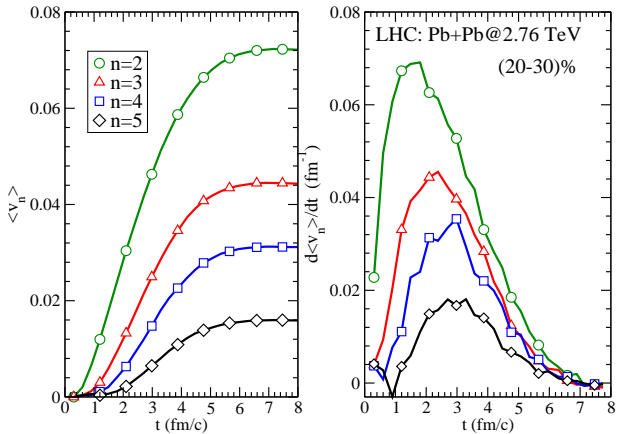}
\end{center}
\caption{Left panel: time evolution of $\langle v_n\rangle$ at mid rapidity respectively for $(20-30)\%$ centrality collisions. Different symbols are for different harmonics.
Right panel: Production rate $\frac{d\langle v_n\rangle}{dt}$ as a function of time at mid rapidity and for the same centrality. These results are for $Pb+Pb$ collision at $\sqrt{s_{NN}}=2.76 \, TeV$.
\label{Fig:dvn_dt}}
\end{figure}
As shown in left panel of Fig.\ref{Fig:dvn_dt} we observe that in the time evolution of the different harmonics $\langle v_n\rangle$ the ordering is present also at late times.
In right panel of Fig.\ref{Fig:dvn_dt} it is shown the production rate for the different harmonics and 
as shown different harmonics have different production rates. In particular, we observe that at very early time the 
second harmonic has a non zero value for $\frac{d\langle v_2\rangle}{dt}\neq 0$ at variance with higher harmonics for 
which $\frac{d\langle v_{n}\rangle}{dt} \approx 0$. 
This different behaviour could be the origin of the stronger correlation between the final elliptic 
flow $v_2$ and its initial eccentricity $\epsilon_2$ that becomes weaker 
between the final $v_n$ and the initial $\epsilon_n$ for higher harmonics ($n > 2$), see Section \ref{section:corr}.

\begin{figure}
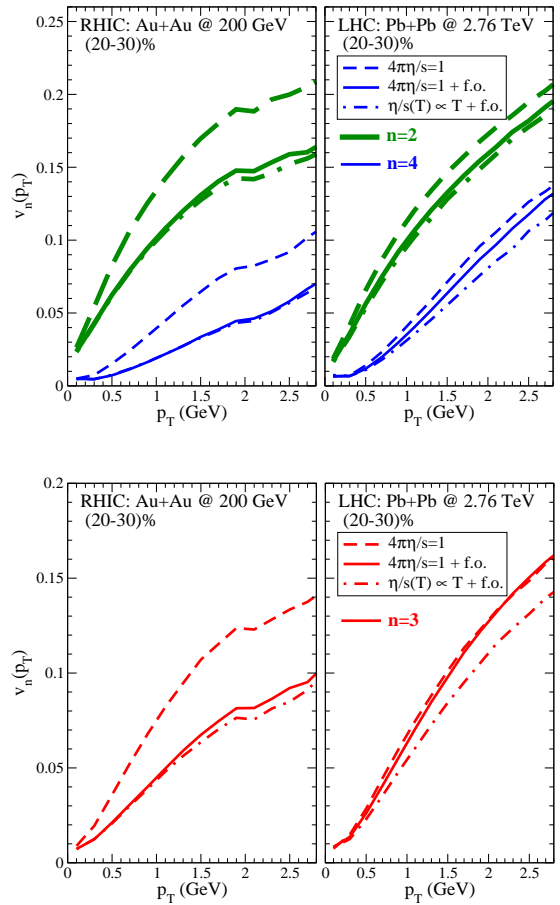

\begin{center}
\includegraphics[width=17pc]{v2n_pt_RHIC_LHC_24.eps}
\end{center}
\hspace{1pc}%
\begin{center}
\includegraphics[width=17pc]{v2n_pt_RHIC_LHC_3.eps}
\end{center}
\caption{Upper panel: differential $v_2(p_T)$ (thick lines) and $v_4(p_T)$ (thin lines) green and blue lines respectively at mid rapidity and for $(20-30) \%$ collision centrality. The comparison is between the two systems: $Au+Au$ at $\sqrt{s}=200 \, GeV$ (left) and $Pb+Pb$ at $\sqrt{s}=2.76 \, TeV$(right). The dashed lines refer to the case with a constant $\eta/s=(4 \pi)^{-1}$ during all the evolution. The solid lines refer to the case with $\eta/s=(4 \pi)^{-1}$ at higher temperature and with an increasing $\eta/s$ ratio at lower temperature while the dot dashed lines to the case with $\eta/s \propto T$ at higher temperature and with an increasing $\eta/s$ ratio at lower temperature. Right panel: differential $v_3(p_T)$ in red lines with the same legend as in the upper panel.
\label{Fig:v2v3v4}}
\end{figure}

Differential flow coefficients $v_n(p_T)$ are observables that carry out more 
information about the fireball created in the heavy ion collisions in particular 
because they are sensitive to the transport properties of the medium like the 
$\eta/s$ ratio. In the following discussion we will study the effect of the $\eta/s$ 
on the build up of the elliptic flow $v_2(p_T)$ and on the high order harmonics $v_3(p_T)$, 
$v_4(p_T)$ and $v_5(p_T)$. With $v_n(p_T)$ we mean the root mean square 
$\sqrt{\langle v_n^2 \rangle}$ as it has been done in experimental 
data using the event plane method.
In the upper panel of Fig.\ref{Fig:v2v3v4} it is shown the elliptic 
flow $v_2(p_T)$ (green thick lines) and the $v_4(p_T)$ (blue thin lines) 
at mid rapidity and for $(20-30) \%$ centrality for both RHIC $Au+Au$ at 
$\sqrt{s} = 200 \,GeV$ (left panel) and LHC $Pb+Pb$ at 
$\sqrt{s}= 2.76\,TeV$ (right panel). In general in agreement with 
what has been obtained in viscous hydrodynamical calculations, the 
increase of the viscosity of the medium has the effect to 
reduce both $v_2$ and $v_4$.

As we can see at RHIC energies comparing the thick dashed lines 
with the solid ones, in the left panel of Fig.\ref{Fig:v2v3v4}, 
the $v_2(p_T)$ is sensitive to the increase of the $\eta/s$ at lower 
temperature close to the cross over region. 
In particular the effect is a reduction of the elliptic flow of 
about $17 \%$. A similar trend were observed for the 4-th harmonic 
$v_4(p_T)$ where we have a reduction due to the increase of $\eta/s$ at 
lower temperature but the effect in this case is 
about a factor two larger then the previous one, i.e. about $30-40 \%$. The different sensitivity to the 
$\eta/s$ can be attributed to their different formation time, 
$t_{v_4}>t_{v_2}$ \cite{Greco:2008fs}.
As shown in Fig.\ref{Fig:vn_t} each harmonics $v_n$ start to develop 
at different times. In particular $v_4$ has its maximum development  
approximatively at $\tau \approx 3 \, fm/c$ while the $v_2$ at $\tau \approx 1.2 \, fm/c$. This means that different harmonics 
probe mainly different temperatures and different value of the $\eta/s$ ratio. 
Assuming that the first few $fm/c$ of the expanding fireball 
are dominated by the 1D longitudinal expansion \cite{Ruggieri:2013bda} where approximatively 
$T(\tau)=T_0 (\tau_0/\tau)^{1/3}$ we have that when $v_4$ has its maximum development at about 
$\tau \approx 3 \, fm/c$ the temperature is $1.3 T_C$ at RHIC and 
$2T_C$ at LHC. In other words this tell us that $v_4$ at 
RHIC energies mainly develops closer to the cross over region
where $\eta/s$ should increase.

On the other hand at LHC energies, left panel of Fig.\ref{Fig:v2v3v4}, 
the scenario is different, the elliptic flow is almost 
unaffected by the increase of $\eta/s$ ratio at low temperature (in the 
hadronic phase) as we can see comparing the green thick dashed line with the 
solid one. Instead we observe that the increase of $\eta/s$ at lower 
temperature has a more sensitive effect on the $v_4(p_T)$ with a reduction 
of about $5 - 10 \%$, see blue solid and dashed lines. Again this different 
sensitivity to the $\eta/s$ in the cross over region between $v_2$ and 
$v_4$ at LHC are consistent with the results obtained at RHIC energies and 
depends on the different formation time of the harmonics
in relation to the initial T of the system. The 
greater sensitivity at RHIC energies of both $v_2$ and $v_4$ to the 
$\eta/s$ at low temperature is related to the different life time of 
the fireball. In fact the life time of the fireball at LHC is greater 
than that at RHIC, 8-10 $fm/c$ at LHC against 4-5 $fm/c$ at RHIC. In 
general this means that at RHIC energies the $v_n$ have not enough time 
to fully develop in the QGP phase. While at LHC energies we have that 
the $v_n$ develops almost completely in the QGP phase and therefore it 
is less sensitive to the dynamics in the cross-over and hadronic region.
This result were firstly found w/o initial state fluctuation in refs.\cite{Plumari_Bari,Plumari:2012xz,Niemi:2012ry}
but remain similar also with fluctuations. The last however allow to study 
for the first time a similar effect also on $v_3(p_T)$.

In Fig.\ref{Fig:v2v3v4} it is shown the effect of an $\eta/s(T)$ in the 
QGP phase. In the comparison between the solid lines 
and the dot dashed ones the only difference is in the linear temperature 
dependence of $\eta/s \propto T$ for $T>T_C$ while at lower temperature 
we have the same dependence (see dot dashed lines in Fig.\ref{Fig:etas_T}). 
As we can see the $v_4$ at LHC is sensitive to the 
change of $\eta/s$ at higher temperature while at RHIC energies the $v_4$ is 
completely unaffected by this change.
In the lower panel of Fig.\ref{Fig:v2v3v4} it is shown the triangular 
flow $v_3(p_T)$ (red lines) at mid rapidity for $(20-30) \%$ 
centrality and for both RHIC $Au+Au$ at $\sqrt{s} = 200 \,GeV$ 
(left panel) and LHC $Pb+Pb$ at $\sqrt{s}= 2.76\,TeV$ (right panel). 
In agreement with what has been obtained for the even harmonics $v_2$ 
and $v_4$, we observe at RHIC energies a reduction of $v_3(p_T)$ due 
to the increase of the $\eta/s$ at low temperature with a reduction 
of about $25 \%$, while at LHC it is almost insensitive to the change of 
$\eta/s$ in the cross over region. 
However we observe that at LHC the third and fourth harmonics are more 
sensitive to the change of $\eta/s(T)$ with respect to the elliptic flow 
with a deviation of about $10\%$ for $v_3$ and $v_4$ against a less 
$5\%$ for $v_2$. 
Still it has be noted that such a sensitivity is quite small to hoping 
a determination of the T dependence of $\eta/s$ from the $v_n(p_T)$.
\begin{figure}
\begin{center}
\includegraphics[width=20pc]{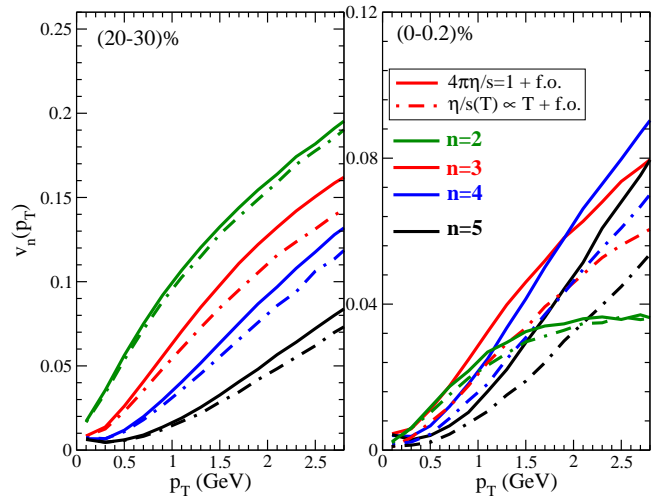}
\end{center}
\caption{
Comparison between $v_{n}(p_{T})$ for mid-peripheral (left panel) and central (right panel) collision. Different colors refer to different harmonics while solid lines correspond to $4\pi\eta/s=1$ in QGP phase and f.o. and dot dashed lines to $\eta/s \propto T$ in the QGP phase and f.o.
\label{Fig:v2v3v4v5_b0_b7.8}}
\end{figure}

\begin{figure}
\begin{center}
\includegraphics[width=18pc]{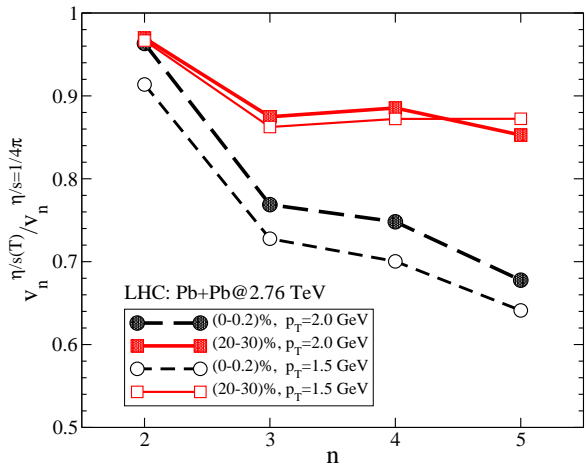}
\end{center}
\caption{Ratio between $v_{n}(p_{T})$ for two different parametrizations of $\eta/s$ as a function of the order 
of the harmonic $n$ and for $p_T=1.5 GeV$ (open symbols) and $p_T=2 GeV$ (full symbols). 
Solid lines refer to mid peripheral collisions while dashed lines to ultra-central collisions.
\label{Fig:ratio}}
\end{figure}

Very recently it has been possible to access also experimentally to the 
ultra-central collisions. The ultra-central collisions are interesting
because the initial $\epsilon_n$ come completely from the fluctuations 
in the initial geometry rather than by global geometric overlap region.
In Fig.\ref{Fig:v2v3v4v5_b0_b7.8} it is shown the comparison of $v_{n}(p_T)$ 
produced in  $Pb+Pb$ at $\sqrt{s}=2.76 \, TeV$ collisions for different centralities: 
left panel for mid-peripheral collisions and right panel for central collisions. 
Different colors are for different harmonics. Solid lines refer to the 
case with $\eta/s=1/(4 \pi)$ in the QGP phase and the increase at low 
temperature as shown in Fig.\ref{Fig:etas_T} by red solid lines while the dot-dashed 
lines refer to the case with $\eta/s \propto T$ in the QGP phase and the increase at low 
temperature as shown in Fig.\ref{Fig:etas_T} by blue dot dashed lines.
From the comparison we observe that at low $p_T$ both centralities the 
$v_{n}(p_T)$ are much flatter for larger $n$.
This results is in agreement with that obtained in hydrodynamic 
calculations where $v_{n}(p_T) \propto p_{T}^{n}$ \cite{Hatta:2014jva}.
On the other hand at hight $p_T$ for ultra-central collisions we observe that 
the elliptic flow $v_{2}(p_T)$ shows a saturation while for $n \ge 3$  
$v_{n}(p_T)$ increase linearly with $p_T$. This is in qualitative agreement with what has been 
observed experimentally, but a quantitative comparison would 
require the inclusion of hadronization that however would not affect the sensitivity to $\eta/s(T)$.
In particular the sensitivity to the value of $\eta/s$ in the QGP phase increase 
with the increasing the order of the harmonics $n$ in agreement with the fact that 
viscous corrections to $v_n(p_T)$ increase with the harmonics \cite{Plumari:2015sia}. 
Furthermore we observe that reduction of $v_{n}(p_T)$ due to the increase of $\eta/s$ in the QGP phase 
(dot dashed lines) is strongly enhanced for ultra-central collisions. As shown in Fig.\ref{Fig:ratio}
for $n \ge 3$ the reduction for central collisions is about $30 - 35 \%$ against 
a reduction of about $10 \%$ for mid peripheral collisions. 
It is indeed remarkable that a $30 \%$ effect is determined by a slowly linear rising of $\eta/s$ with T 
as the one considered and depicted in Fig.\ref{Fig:etas_T}. 
In particular in central collisions higher harmonics acquire a larger sensitivity to the value 
of the viscosity in the QGP phase.
Therefore our study suggest that to have information about $\eta/s(T)$ one should focus on 
ultra-central collisions. This point is further strenghtened by the study of the correlations 
between $v_n$ and the initial eccentricities $\epsilon_n$ that we discuss in the next section.


\section{Correlations between $v_n$ and $\epsilon_n$}\label{section:corr}

\begin{figure*}
\includegraphics[width=12pc]{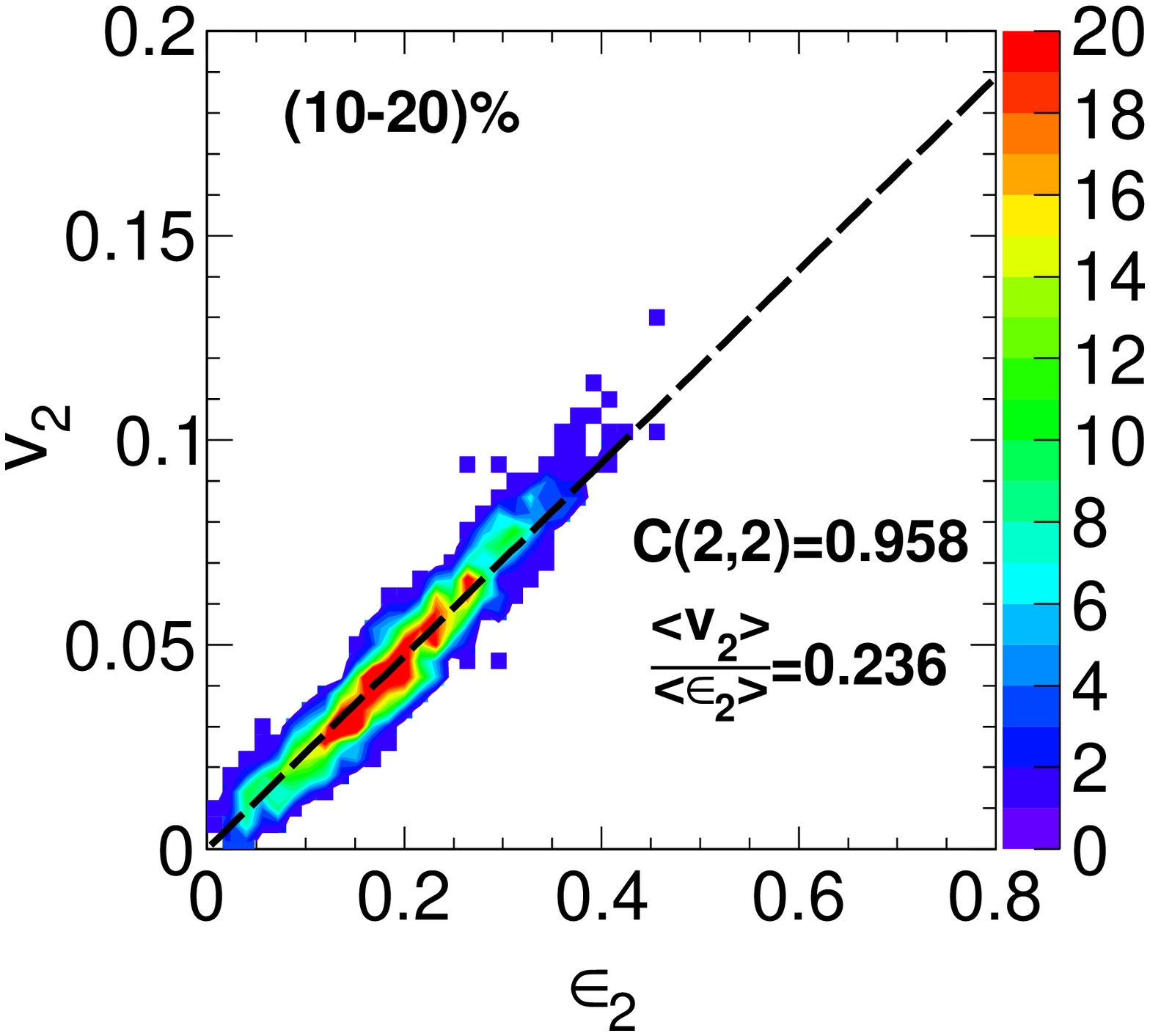}
\includegraphics[width=12pc]{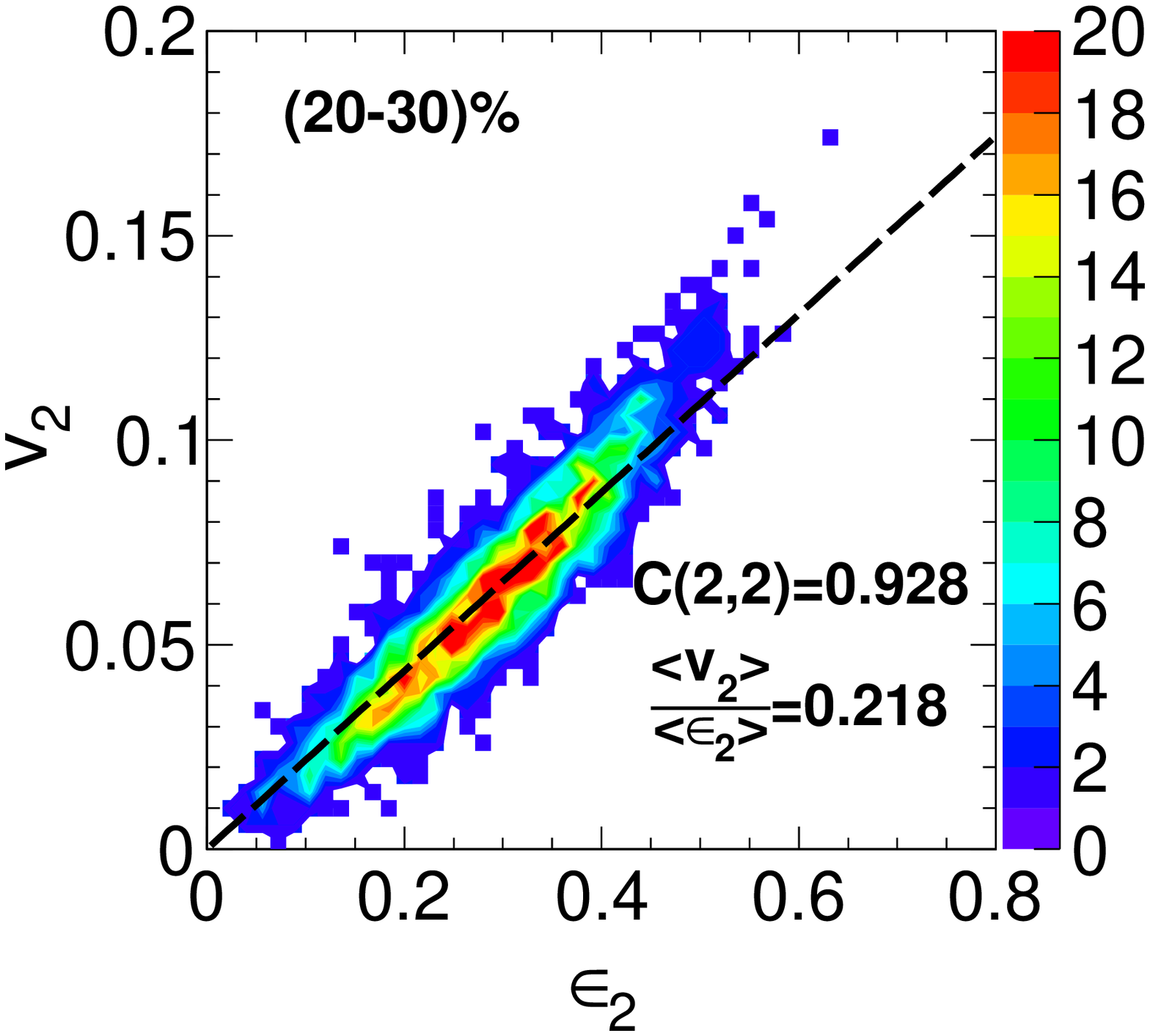}
\includegraphics[width=12pc]{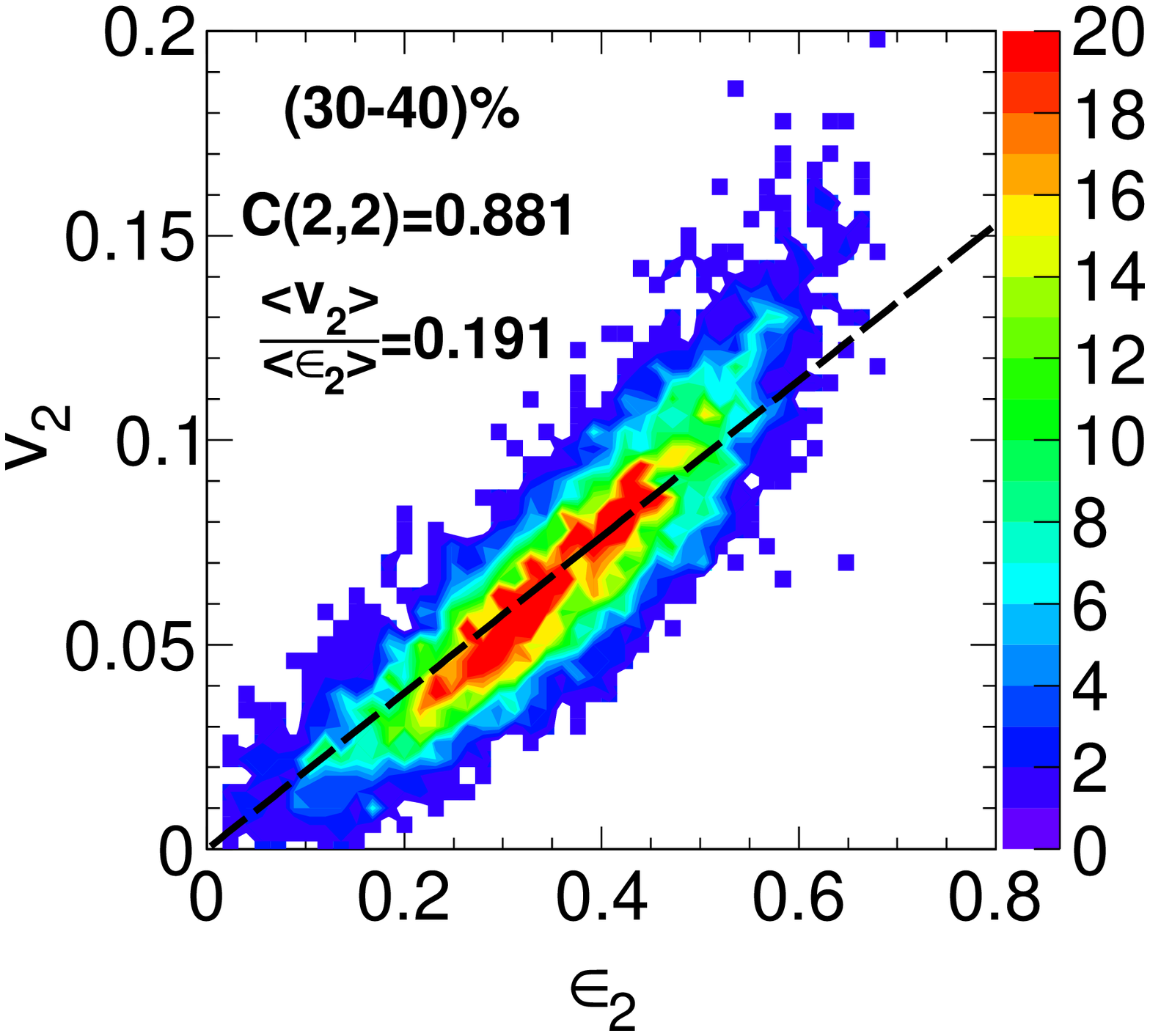}
\includegraphics[width=12pc]{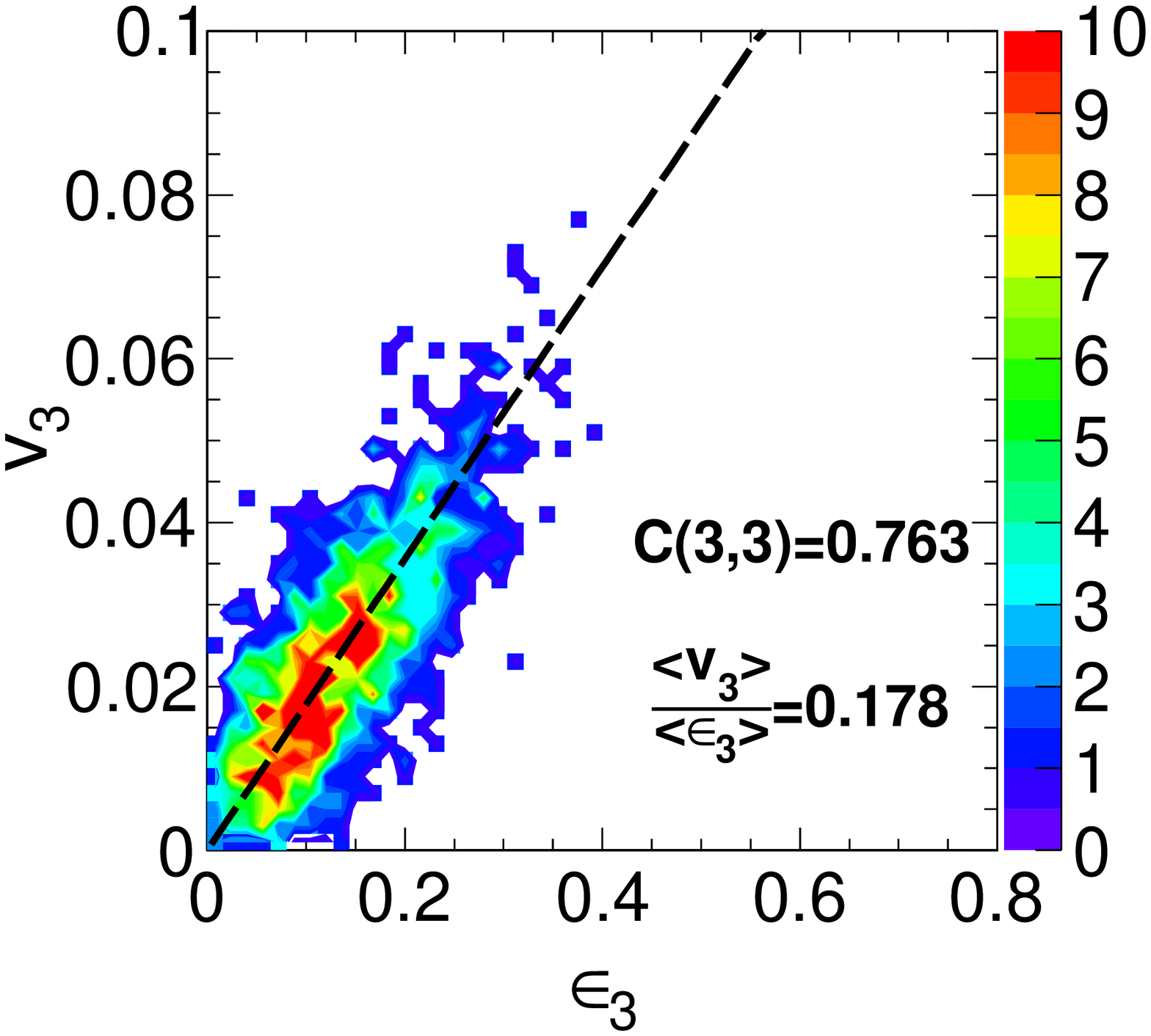}
\includegraphics[width=12pc]{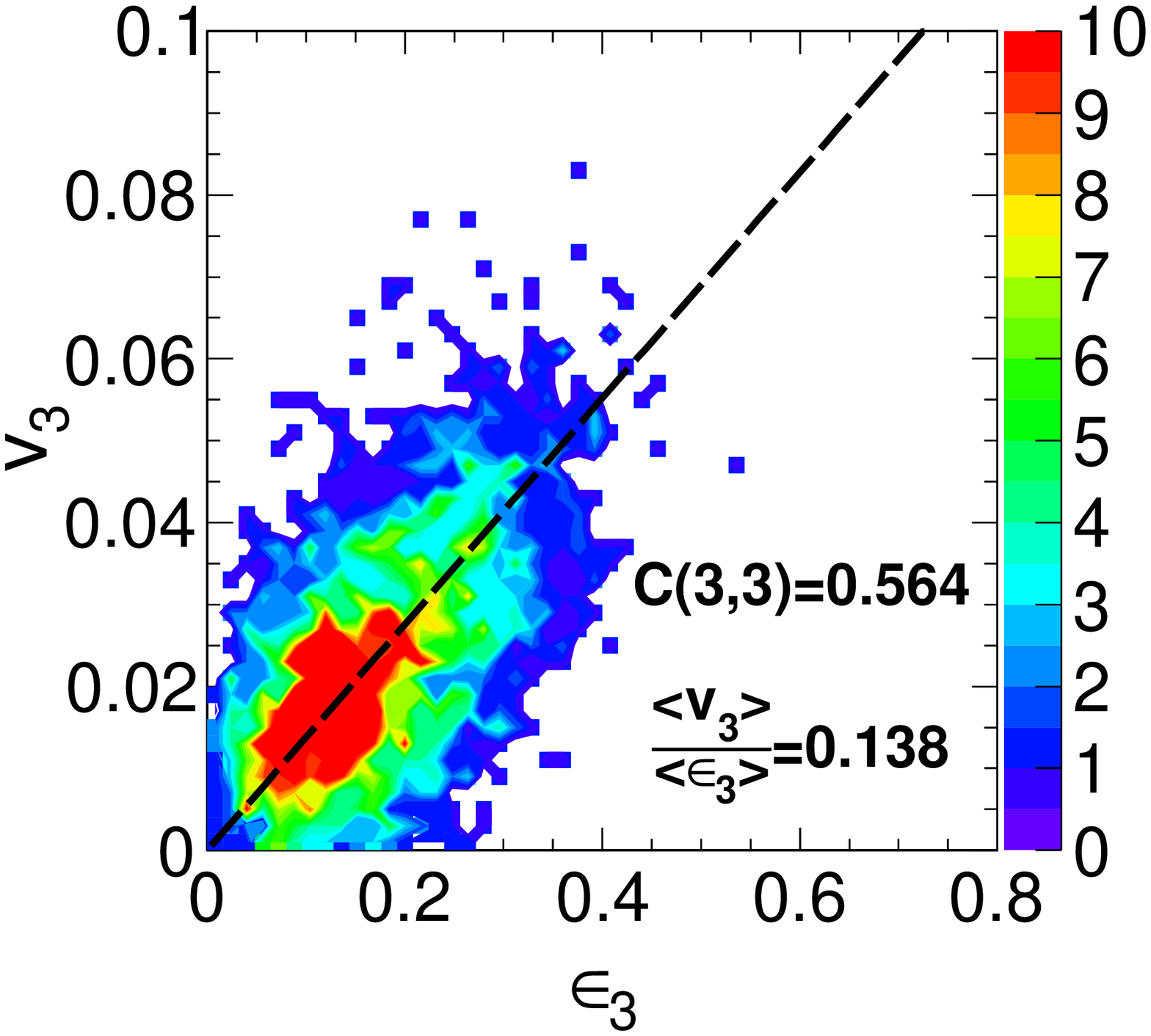}
\includegraphics[width=12pc]{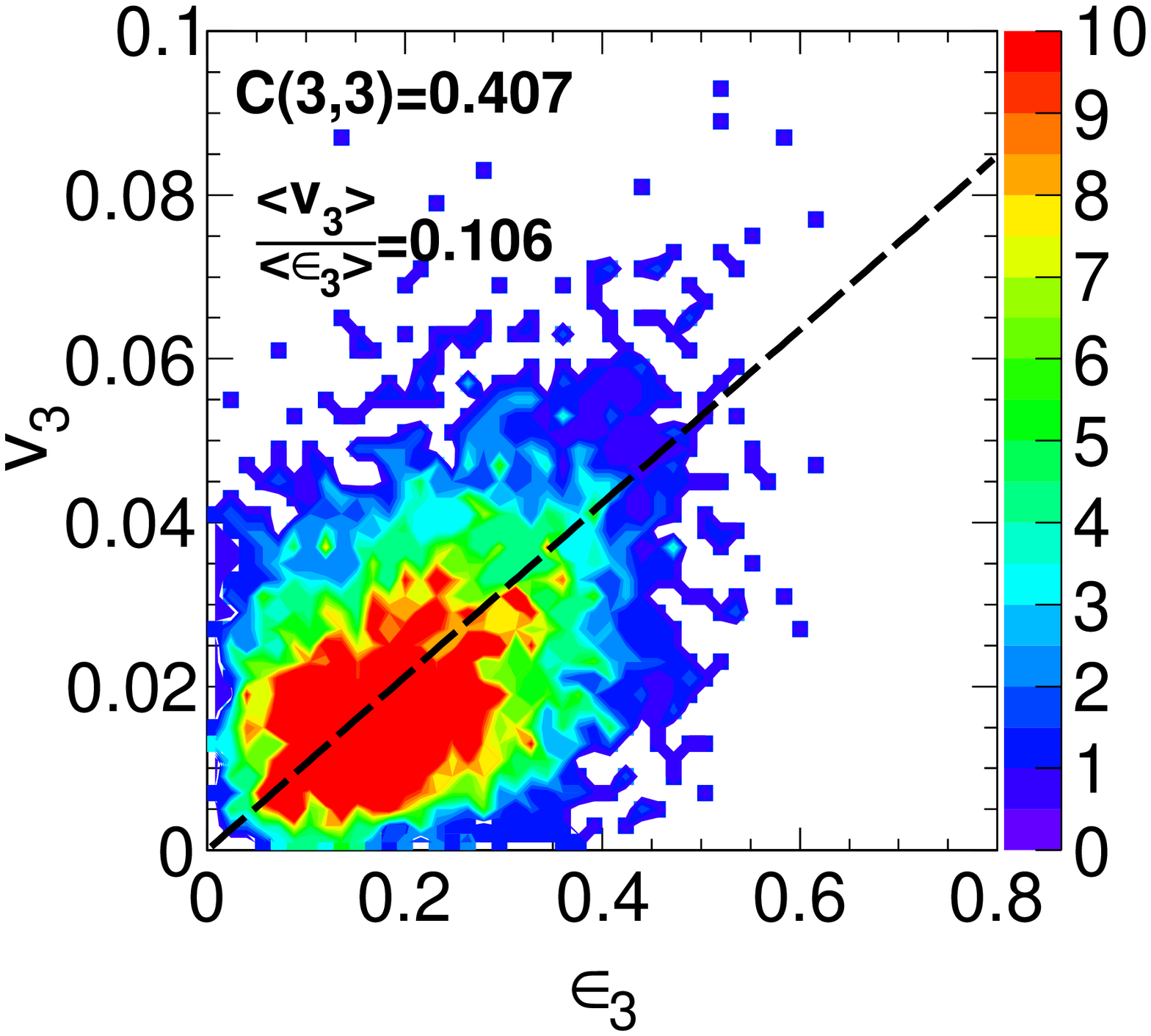}
\includegraphics[width=12pc]{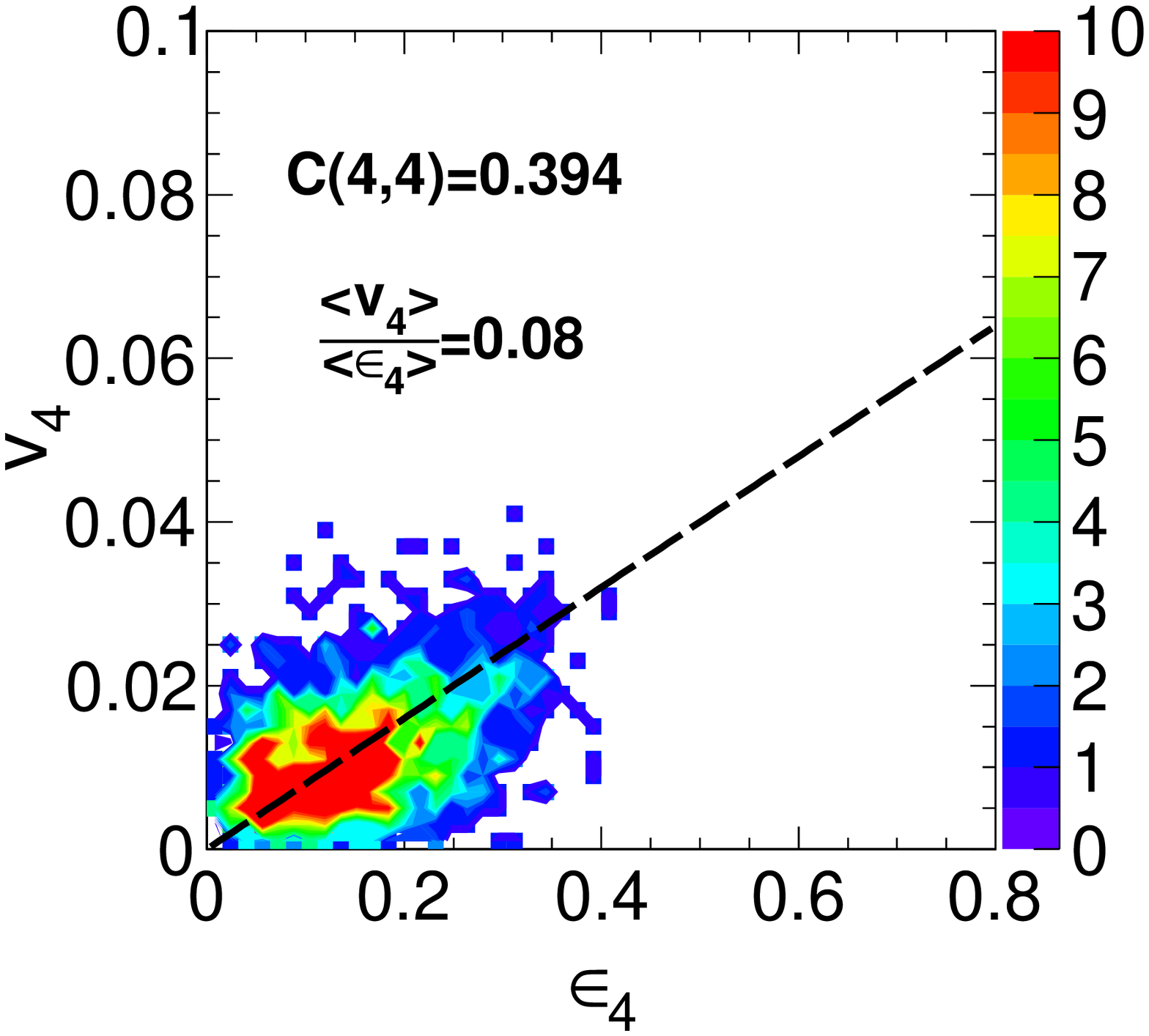}
\includegraphics[width=12pc]{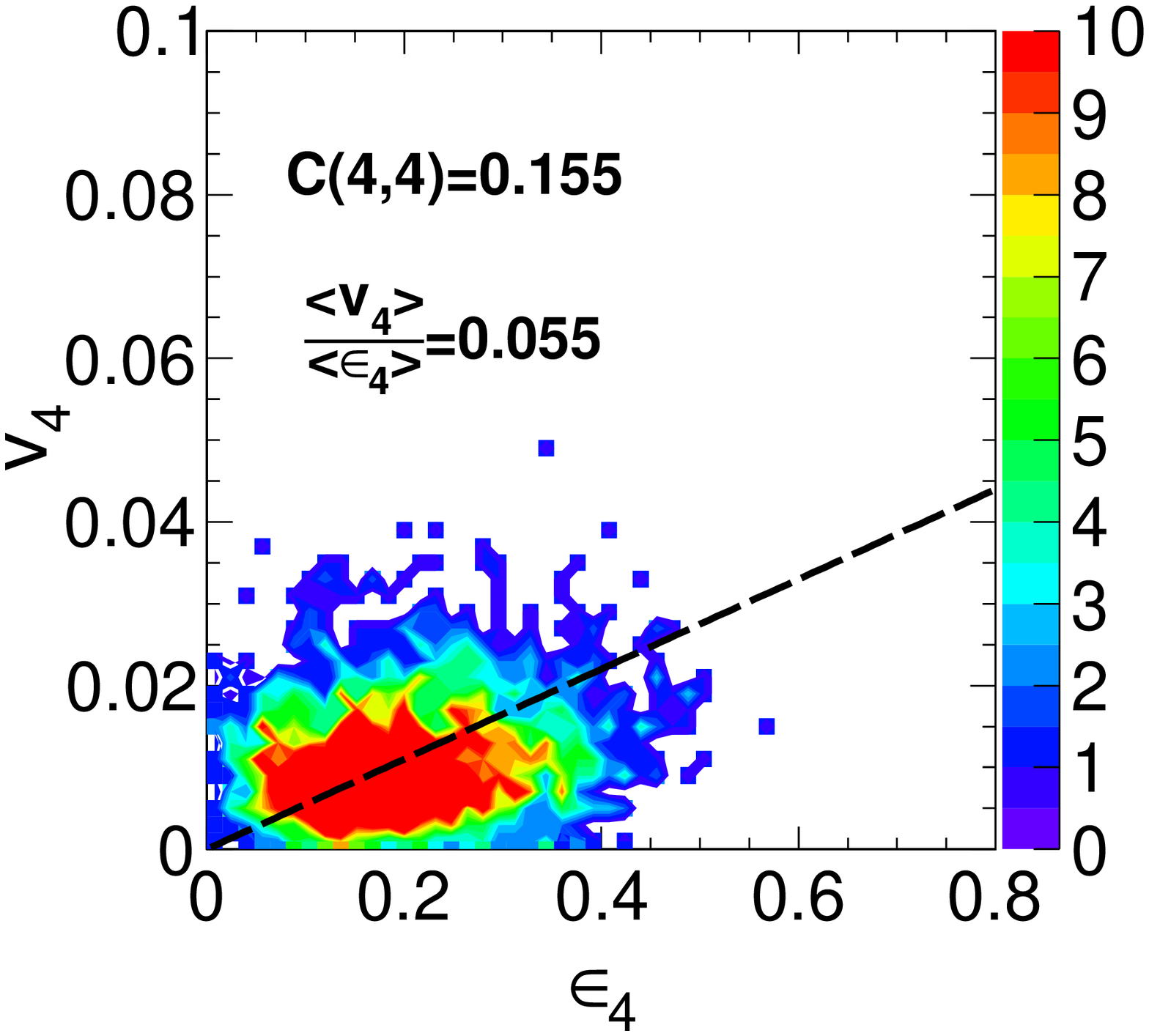}
\includegraphics[width=12pc]{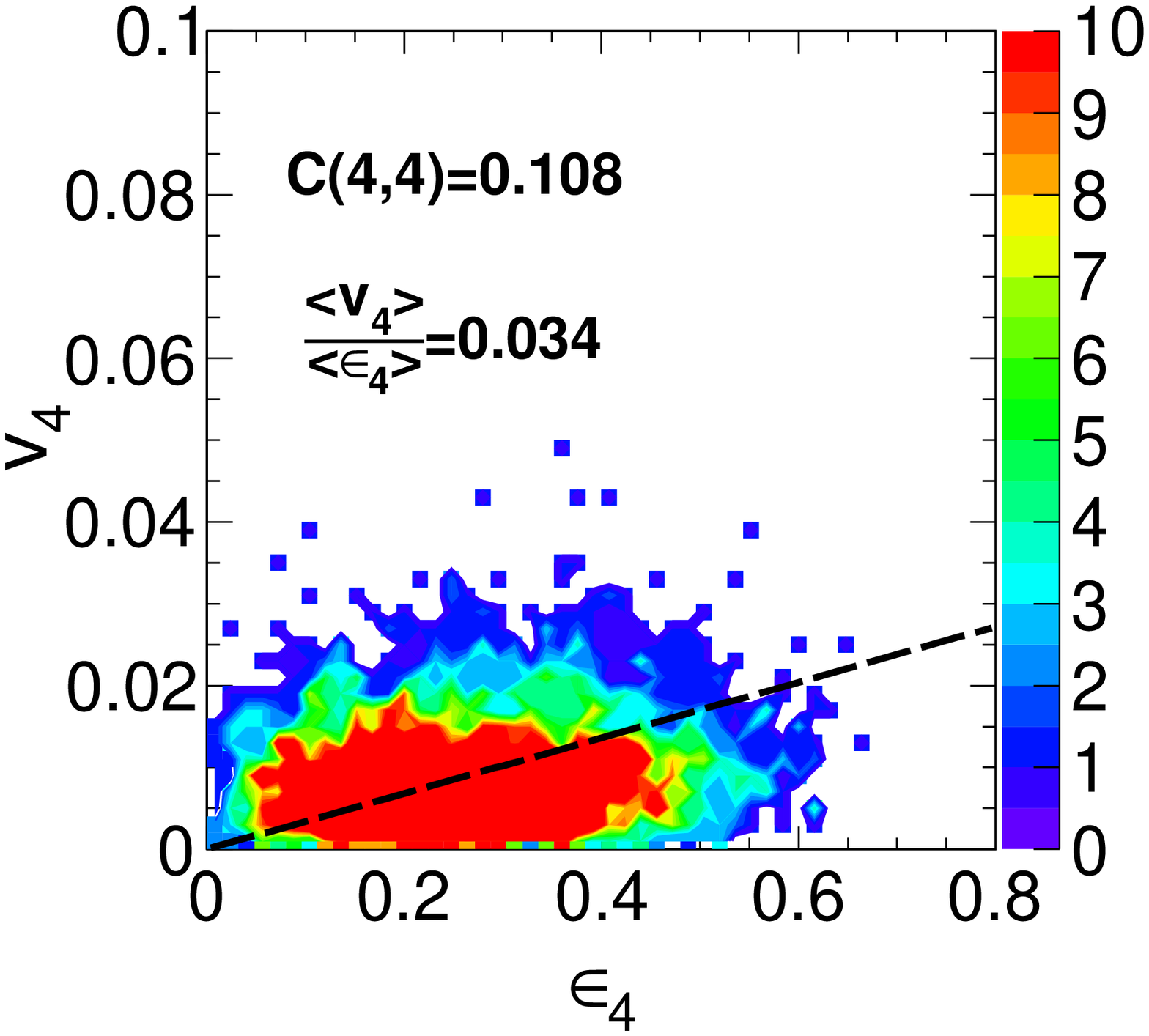}
\caption{ $\epsilon_n$ and $v_n$ for $Au+Au$ collisions at $\sqrt{s_{NN}}=200 \, GeV$ and for three different centrality class.
Upper panel: $\epsilon_2$ and $v_2$ for $(10-20)\%$, $(20-30)\%$ and $(30-40)\%$ from left to right.
Middle panel: $\epsilon_3$ and $v_3$ for the same centralities. Finally in the lower panel $\epsilon_4$ and $v_4$.
In these calculations we have fixed $\eta/s=1/(4\pi)$ for hight temperature and the kinetic f.o. at lower temperature (see solid line in Fig.\ref{Fig:etas_T}).
\label{Fig:corr}}
\end{figure*}

In recent years, the correlation between integrated
$v_2$ and high order harmonics $v_3, \, v_4$ with the 
initial asymmetry in coordinate space $\epsilon_2, \epsilon_3$ 
and $\epsilon_4$ have been studied in the event-by-event ideal and viscous 
hydrodynamics framework \cite{Gardim:2011xv,Chaudhuri:2012mr,Niemi:2012aj}. 
In general it has been shown that the elliptic flow is strongly correlated 
with initial eccentricity while a weaker correlation has been found 
for higher harmonics $v_3,v_4$ with $\epsilon_3$ and $\epsilon_4$.
One explanation for the weak correlation observed between $v_4$ and $\epsilon_4$ 
is that for final $v_4$ there is also a correlation with the initial $\epsilon_2$. 
In particular in \cite{Gardim:2011xv} has been shown that it possible to have a good 
linear correlation between $v_4$ and a linear combination 
of the initial $\epsilon_2$ and $\epsilon_4$.

In this section we discuss these correlations within an event by event transport 
approach with initial state fluctuations. 
A measure of the linear correlation is given by the correlation coefficient 
$C(n,m)$ given by the following expression:
\begin{equation}
C(n,m)=\frac{\sum_{i}(\epsilon_n^i-\langle \epsilon_n \rangle)(v_m^i-\langle v_m \rangle)}{\sqrt{\sum_{i}(\epsilon_n^i-\langle \epsilon_n \rangle)^2\sum_{i}(v_m^i-\langle v_m \rangle)^2}}
\end{equation}
where $\epsilon_n^i$ and $v_m^i$ are the values of $\epsilon_n$ and $v_m$ corresponding 
to the given event $i$ and evaluated according Eq.s (\ref{Eq:eps_n}) and (\ref{Eq:v_n}). 
$C(n,m) \approx 1$ corresponds to a strong linear correlation between the initial 
$\epsilon_n$ and the final $v_m$.

The results shown in this section have been obtained with 
$N_{event}=1000$ events for each centrality class a total number of test particle per event $N_{test}=2 \cdot 10^{6}$.
In Fig.\ref{Fig:corr}, it is shown the two-dimensional plots of the integrated 
flow coefficients $v_n$ as a function of the corresponding 
initial $\epsilon_n$ for each event.
The results shown are for $Au+Au$ collisions at $\sqrt{s_{NN}}=200 \, GeV$ 
and for three different centralities $(10-20)\%$, $(20-30)\%$ and $(30-40)\%$.
The viscosity has been fixed to $4\pi\eta/s=1$ plus a kinetic f.o. realized by the increase 
in $\eta/s(T)$ as in Fig.\ref{Fig:etas_T}.
As shown in the upper panel we observe a stronger linear correlation between $\epsilon_2$ and $v_2$ 
for mid central collisions with a linear correlation coefficient that shows a monotonic 
behaviour with the collision centrality from $C(2,2) \approx 0.96$ for $(10-20)\%$ to 
$C(2,2) \approx 0.89$ for $(30-40)\%$.
Qualitatively the results are in agreement with the one obtained within a 2+1D viscous 
hydrodynamics, see \cite{Niemi:2012aj}. 
In general we observe a slightly smaller degree of correlation probably induced by the 
fact that we simulate a 3+1D expansion that can be expected to contribute to the decorrelation. 
In the middle and lower panel of Fig.\ref{Fig:corr} we have shown similar plots for the third 
and fourth harmonics. We observe again a reduction of the correlation coefficient with the 
centrality of the collision similarly to $v_2$ and $\epsilon_2$.
We obtain that the correlation between $\epsilon_3$ and $v_3$ for all the collision centralities 
is weaker with respect to that obtained for the elliptic flow. 
Furthermore the fourth harmonic flow $v_4$ shows a weak correlation 
with the initial $\epsilon_4$ in particular for mid-peripheral collisions where the linear 
correlation coefficient is quite weak $C(4,4) < 0.3$.
Furthermore we observe that the $\langle v_n \rangle/\epsilon_n$ ratio (see dashed lines in Fig.\ref{Fig:corr}) decreases when decreases 
the correlation coefficient $C(n,n)$, i.e. for more peripheral collisions.

\begin{figure}
\begin{center}
\includegraphics[width=21pc]{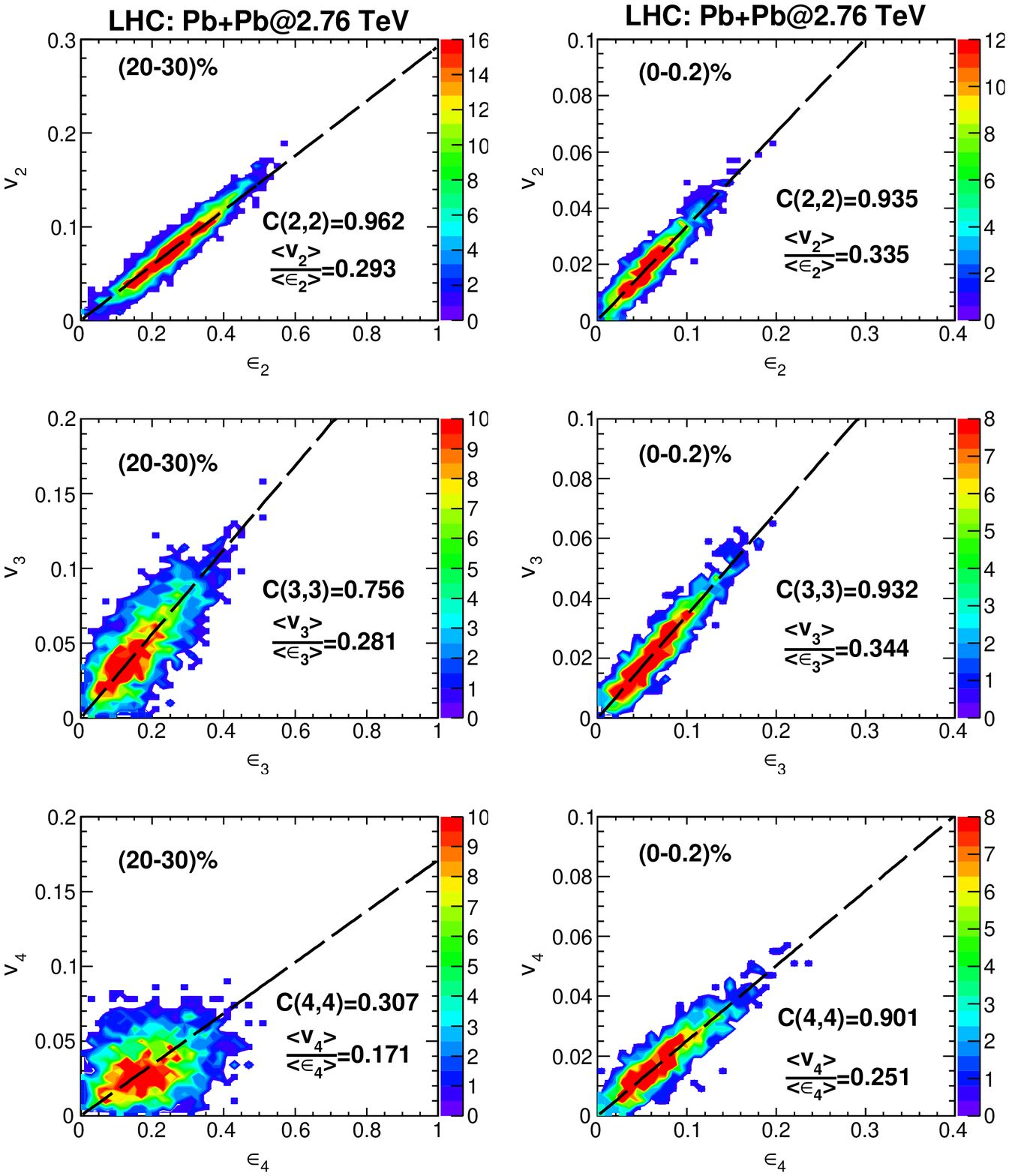}
\end{center}
\caption{
$\epsilon_n$ and $v_n$ for $Pb+Pb$ collisions at $\sqrt{s_{NN}}=2.76 \, TeV$ and for $(20-30)\%$ and $(0-0.2)\%$ centrality cut respectively righ and left panel. In these calculations we have fixed $\eta/s=1/(4\pi)$ for hight temperature and the kinetic f.o. at lower temperature (see solid line in Fig.\ref{Fig:etas_T}).
\label{Fig:compare_RHIC_LHC}}
\end{figure}
A similar behaviour for the linear correlation coefficient $C(n,n)$ is observed at LHC energies 
for $Pb+Pb$ collisions at $\sqrt{s_{NN}}=2.76 \, TeV$. In Fig.\ref{Fig:compare_RHIC_LHC} it is 
shown the comparison between ultra-central and mid-peripheral collision at LHC energies at $\sqrt{s_{NN}}=2.76 \, TeV$.
In ultra-central collisions the $\langle v_n \rangle$ are more correlated to the initial $\epsilon_n$ than at peripheral 
collisions and very interesting differences emerge looking also at higher harmonics.

\begin{figure}
\begin{center}
\includegraphics[width=19pc]{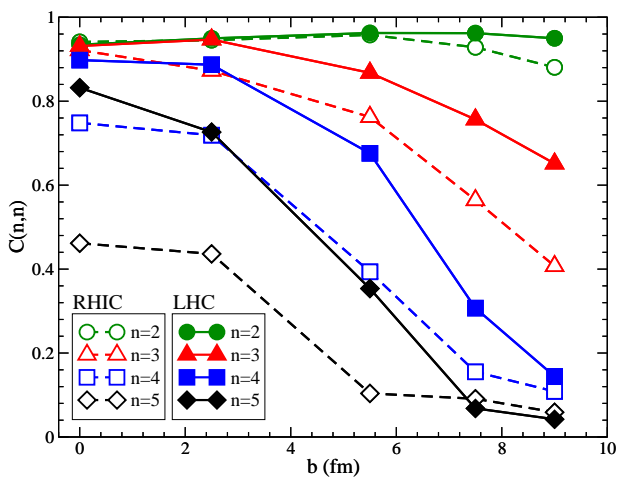}
\end{center}
\caption{Correlation coefficient $C(n,n)$ as a function of the impact parameter b. Different symbols refer to different harmonics $n$. In particular circles, triangles, squares and diamonds refer to $n=2,3,4$ and $5$ respectively. The solid lines correspond to $Au+Au$ collisions at $\sqrt{s_{NN}}=200 \, GeV$ while dashed lines to the system  $Pb+Pb$ at $\sqrt{s_{NN}}=2.76 \, TeV$.
\label{Fig:corr_RHIC_LHC}}
\end{figure}

In order to better visualize and discuss such differences we have plot in Fig.\ref{Fig:corr_RHIC_LHC} the 
$C(n,n)$ as a function of the impact parameter for both RHIC (dashed lines) and LHC energies (solid lines).
As shown the linear correlation coefficient is a decreasing function of the impact parameter 
for all the harmonics. 
However, as we can see comparing the dashed and solid lines show, at 
LHC energies there is a stronger correlation between $\epsilon_n$ and $v_n$ for all $n$ with 
respect to RHIC energies.  
We observe that $v_2$ and $\epsilon_2$ have the same degree of correlation for both RHIC 
and LHC energies while a lower degree of correlation it is shown for higher harmonics 
$n=3, \, 4$ and $n=5$. 
More interesting is the fact that for ultra-central collisions at LHC 
the linear correlation coefficient $C(n,n)$ remains above $0.9$ for $n=2,3,4$
and even the $n=5$ shows large $C(n,n)=0.85$. This can be visualized also in the right 
panel of Fig.\ref{Fig:compare_RHIC_LHC} where the ($v_n$,$\epsilon_n$) correlation plot 
is shown in mid-peripheral ($20-30\%$) collisions (left panel) and in ultra-central 
collisions ($0-0.2\%$) (right panels).

The strong correlation observed for ultra-central collisions means that the value 
obtained for $\langle v_n \rangle$ and its dependence with the harmonics $n$ for those 
collisions is strongly related to the value of the initial asymmetry measure $\epsilon_n$. 
In particular this could imply that the structure of the $v_n(p_T)$ at LHC where $C(n,n) \approx 1$ carry out information
about the initial geometry of the fluctuations.
This joined to the observation that for ultra-central collisions the sensitivity of $v_{n}$ to $\eta/s$ is increased by 
about a factor of 2-3 strongly suggests to focus the experimental efforts at LHC highest energy and ultra-central collisions.

To study the effect of the viscosity and its possible temperature dependence on 
the correlation we have studied how change the correlation coefficient with 
the different parametrizations for $\eta/s$.
In Table \ref{table:corr} we show the results for $C(n,n)$ for the two energies RHIC and LHC for 
$(20-30)\%$ centrality class.
In general for this centrality we observe that at LHC energies and for all the 
viscosities considered the degree of correlation between $\epsilon_n$ and $v_n$ is 
greater than the one at RHIC energies. Moreover we obtain that at LHC the correlation 
coefficient is not sensitive to the change of the viscosity both at low and high 
temperature. A slight different behaviour we have at RHIC energies where 
the effect of the kinetic freeze out is to reduce the degree of correlation 
between the initial $\epsilon_n$ and the final $v_n$.
\begin{table}[t]
\caption{Linear correlation coefficient $C(n,n)$ for RHIC and LHC energies and for different temperature parametritazion of $\eta/s$. These results are for $(20-30)\%$ centrality class.}
\begin{center}
\begin{tabular}{c|c|c|c|c}
$C(n,n)$ & $n$ & $4 \pi \eta/s=1$ & $4 \pi \eta/s=1$ + f.o. & $\eta/s \propto T$ + f.o.\\
\hline
     & 2& 0.95& 0.94& 0.93 \\
RHIC & 3& 0.70& 0.58& 0.65 \\
     & 4& 0.30& 0.28& 0.31 \\
\hline
     & 2& 0.96& 0.96& 0.96\\
LHC  & 3& 0.78& 0.78& 0.74\\
     & 4& 0.39& 0.38& 0.38
\end{tabular}
\end{center}
\label{table:corr}
\end{table}
Furthermore, we have computed the non diagonal components for the linear correlation coefficient $C(n,m)$. 
We found that $C(2,3) \approx 0.$  and $ C(3,4) \approx 0$ for all the range of centralities explored 
which means that there is no linear correlation between $v_2$ and $\epsilon_3$
and $v_3$ and $\epsilon_4$. A different behaviour we observe for $C(4,2)$ 
which is seen to be an increasing function with the centrality $C(4,2) \approx 0.02$ 
for central collision ($b=0 \, fm$) and about $C(4,2)\approx 0.23$ at $b=7.8 \, fm$. This means that in 
more peripheral collisions the 4th harmonic $v_4 $  has some contamination of $\epsilon_2$ and it is not driven 
only by $\epsilon_4$ as already suggested in \cite{Gardim:2011xv}.

\begin{figure}
\begin{center}
\includegraphics[width=19pc]{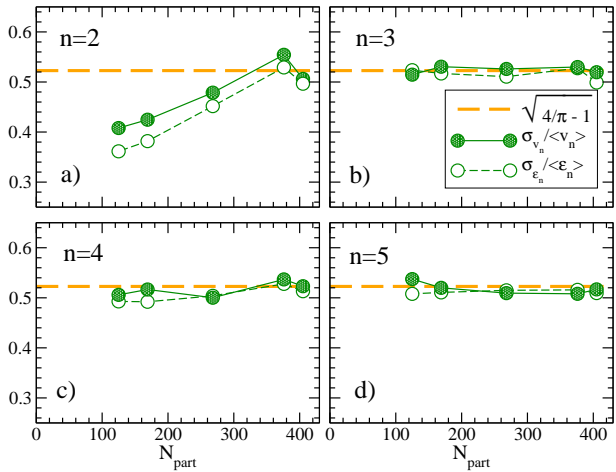}
\end{center}
\caption{From panel a) to d) $\sigma_{v_{n}}/\langle v_n \rangle$ (full symbols) and $\sigma_{\epsilon_{n}}/\langle \epsilon_n \rangle$ (open symbols) as a function of $\langle N_{part} \rangle$ respectively for $n=2,3,4$ and $5$. The dashed lines indicate the value $\sqrt{4/\pi -1}$ expected for a 2D Gaussian distribution. These results are for $Pb+Pb$ collisions at $\sqrt{s_{NN}}=2.76 \, TeV$.
\label{Fig:sigma_vn}}
\end{figure}

Some interesting properties of the $v_n$ distributions can be inferred by studying the 
centrality dependence of the relative fluctuations $\sigma_{v_{n}}/\langle v_n \rangle$.
In Fig.\ref{Fig:sigma_vn}a) it is shown the $\langle N_{part} \rangle$ dependence of the ratios 
$\sigma_{v_{n}}/\langle v_n \rangle$ and $\sigma_{\epsilon_{n}}/\langle \epsilon_n \rangle$ where $\sigma_{v_{n}}$ 
and $\sigma_{\epsilon_{n}}$ are the standard deviation respectively for $v_n$ and $\epsilon_n$. As shown 
for $n=2$ we observe a strong dependence of the relative fluctuations with the centrality of the collision 
with $\sigma_{v_{2}}/\langle v_2 \rangle \approx 0.4$ for $\langle N_{part} \rangle \approx 130$. 
For more central collisions this ratio approaches the value expected for a 2D Gaussian distribution where 
$\sigma_{v_{n}}/\langle v_n \rangle = \sqrt{4/\pi -1} \approx 0.523$ \cite{Aad:2013xma}, 
shown by dashed line in Fig.\ref{Fig:sigma_vn}.
For higher harmonic $n=3,4$ and $5$ as shown from Fig.\ref{Fig:sigma_vn}b) to Fig.\ref{Fig:sigma_vn}d) 
the values of $\sigma_{v_{n}}/\langle v_n \rangle$ are approximatively the same of the ones of the initial geometry with  
$\sigma_{v_{n}}/\langle v_n \rangle \approx \sigma_{\epsilon_{n}}/\langle \epsilon_n \rangle$ and they are almost independent 
of the collision centrality and for all the centralities studied they are very close to the the value $\sqrt{4/\pi-1}$ 
shown by the dashed lines. 
These results imply that the distributions of $v_3,\, v_4$ and $v_5$ for all the centrality range studied 
are consistent with the fluctuation-only scenario discussed in \cite{Aad:2013xma} and these fluctuations are 
related to the fluctuations of the initial geometry. On the other hand, the distribution of $v_2$ is close to this 
limit for most central collisions while for mid-peripheral collisions there is a contribution coming from the global 
average geometry. 

\section{Conclusions}
Using an event-by-event transport approach we have investigated
the build up of the anisotropic flows $v_{n}(p_T)$ for $n=2,3,4$ and $5$.
In particular we have studied the effect of $\eta/s$ ratio on $v_{n}(p_T)$ 
for two different beam energies: at RHIC for $Au+Au$ collisions at $\sqrt{s}=200 \, GeV$ 
and at LHC for $Pb+Pb$ collisions at $\sqrt{s}=2.76 \, TeV$.
We have found that at RHIC the $v_n(p_T)$ 
are more affected by the value of $\eta/s$ at low temperature ($T<1.2 T_C$) and the  
sensitivity increases with the order of the harmonics.
At LHC we get a different effect, all the $v_n(p_T)$ develop in the QGP phase at 
and are not affected by the value of $\eta/s$ in the cross-over region.
However the sensitivity to the T dependence of the $\eta/s$ is quite weak, more 
specifically a constant $\eta/s=0.08$ or an $\eta/s \propto T$ induce differences 
in the $v_2$ of at most a $5 \%$ and of about a $10 \%$ in $v_3, v_4, v_5$.
The novel result from our analysis is that such a scenario changes for ultra-central collisions 
where found an enhancement of the sensitivity of the $v_n(p_T)$ that for $n=3,4,5$
reaches about a $30 \%$.
We have also studied the correlation between the initial asymmetry in coordinate space,
measured by $\epsilon_n$, and the final asymmetry in momentum space given by $\langle v_n \rangle$.
We have found that larger is the collision energy larger is the degree of correlation between $\epsilon_n$
and $\langle v_n \rangle$. At LHC there is significantly more correlation than at RHIC.
For both collision energies considered and in all the range of impact parameter studied the $v_2$
is strongly correlated with the $\epsilon_2$ with the linear correlation coefficient $C(2,2) \approx 0.95$. 
The degree of correlation between $\epsilon_n$ and the corresponding $\langle v_n \rangle$ decrease for 
higher harmonics. 
Moreover, in ultra-central collisions we found that $C(n,n) > 0.9$ for $n=2,3$ and 
$4$ which imply that the $v_n \propto \epsilon_n$ and they carry out the information about the initial 
geometry of the fireball. 
These results joined with the fact that in ultra central collisions the $v_n(p_T)$ have a large sensitivity 
to the $\eta/s$ ratio strongly suggest to focus the experimental effort to these collision centrality where it is possible 
to get better constraint on the value of $\eta/s$ in the QGP phase and having a new insight on the initial state fluctuations.

\section{Acknowledgments}
V.Greco, S. Plumari, F. Scardina and G.L. Guardo 
acknowledge the support of the ERC-StG Grant
under the QGPDyn project.

\end{document}